\documentclass[pra, 12pt, a4paper, floatfix]{revtex4}

\usepackage{chemformula} 
\usepackage[T1]{fontenc} 
\usepackage{amssymb}
\usepackage{amsmath}

\usepackage{hyperref}

\newcommand{\vect}[1]{\boldsymbol{#1}}

\newcommand{\op}[1]{\hat{#1}}

\newcommand{\crop}[2]{\op{#1} _{#2} ^{\dagger}}

\begin{document}

\author{Guan-Hao Peng}
\affiliation{Department of Electrophysics, National Yang Ming Chiao Tung University, Hsinchu 300, Taiwan}

\author{Oscar Javier Gomez Sanchez}
\affiliation{Department of Electrophysics, National Yang Ming Chiao Tung University, Hsinchu 300, Taiwan}

\author{Wei-Hua Li}
\affiliation{Department of Electrophysics, National Yang Ming Chiao Tung University, Hsinchu 300, Taiwan}

\author{Ping-Yuan Lo}
\affiliation{Department of Electrophysics, National Yang Ming Chiao Tung University, Hsinchu 300, Taiwan}

\author{Shun-Jen Cheng}
\affiliation{Department of Electrophysics, National Yang Ming Chiao Tung University, Hsinchu 300, Taiwan}
\email{sjcheng@mail.nctu.edu.tw}

\title{Twisted-light-induced exciton wave packets in transition-metal dichalcogenide monolayers}







\begin{abstract}
We present a comprehensive theoretical investigation of the photo-generated excitons in  transition-metal dichalcogenide monolayers (TMD-ML's) by Laguerre-Gaussian beams, a celebrated kind of  twisted lights (TL's) carrying quantized orbital angular momenta (OAM). We show that the photo-excitation of TL incident to a TMD-ML leads to the formation of spatially localized exciton wave packets, constituted by the superposition of finite-momentum exciton states determined by the intriguing interplay between the multiple degrees of freedom of the optical and excitonic subsystems. 
Consequently, the TL-induced exciton wave packets yield profound directional photo-luminescences whose polar-angle-dependences are encoded by the transferred optical OAM and azimuthal angle part, despite OAM-irrelevant, optically resolves the exchange-split longitudinal and transverse exciton bands.  Interestingly, the application of linearly polarized TL onto a valley-excitonic system mimics an exciton multiplexer allowing for selectively detecting the individual valley-mixed exciton bands, which are normally hardly measured spectrally.
\end{abstract}

\maketitle


\section{Introduction}

With the intriguing electronic and excitonic properties, atomically thin TMD-ML's have drawn vast attention for over a decade and  been well realized nowadays as promising two-dimensional (2D) materials for advanced optoelectronic and valley-based photonic applications. \cite{amani2015near,schaibley2016valleytronics,mueller2018exciton,wang2018colloquium} Because of inherently weak Coulomb screening in the 2D structures, photo-generated electron-hole ({\it e-h}) pairs in a TMD-ML form tightly bound excitons via the enhanced Coulomb attractions, with the exciton binding energy so high as hindreds of meV. \cite{chernikov2014exciton,he2014tightly} It is such tightly bound excitons, rather than the free {\it e-h} pairs, that dictate the major optical features of the atomically thin 2D materials and various extraordinary optical and excitonic phenomena result in the 2D materials, e.g. room-temperature formation of exciton-polariton, \cite{lundt2016room,zhang2018photonic} high-temperature exciton condensation. \cite{wang2019evidence}, ultra-fast excitation energy transfer, \cite{kozawa2016evidence,wu2019ultrafast} superiorly high efficiencies of luminescences \cite{amani2015near} and light harvest, \cite{bernardi2013extraordinary,tsai2014monolayer} and rich exciton fine structures. \cite{chen2018coulomb,liu2020multipath,zhang2017magnetic,molas2017brightening,wang2017inplane,li2019momentum,he2020valley,qiu2015nonanalyticity,brem2020phonon, peng2019distinctive, lo2021fullzone} 

Even more interestingly, an exciton in a TMD-ML possesses multiple degrees of freedom, including spin, valley, and the center-of-mass momentum ($\vect{Q}$) as well.
The spin-valley locking effect in $D_ {3h}$ TMD-ML's enables the selective photo-generatation of exciton in specific $K$ or $K'$ valley and the coherent manipulation of the superposition valley exciton states by controlling the polarization, i.e. optical spin angular momentum (SAM), of the incident light. This sets up the prospect of valley exciton in TMD-ML's for valley-based photonics and quantum technology. \cite{schaibley2016valleytronics,mueller2018exciton}
By contrast to the maturity of the manipulation of the spin-valley-locked degrees of freedom, guiding the motion of the center-of-mass (CoM) of exciton and manipulating the exciton states of different CoM momenta yet remain as a non-trivial task because of the charge neutrality of exciton that hinders the electrically bias-controlled transport. \cite{onga2017exciton,yang2021waveguiding,fedichkin2016room}

The conventional light sources used for the photo-generation of excitons are commonly based on non-structured laser beams that carry the optical SAM only.
In fact, a light can be spatially structured to acquire additional degrees of freedom, e. g. optical OAM. Such structured lights with quantized OAM, also referred to as twisted light (TL), was first predicted by Allen {\it et al.} in the early 90s, \cite{allen1992orbital} and soon later realized experimentally by He {\it et al.}. \cite{he1995direct} Over the past three decades, such new states of photons have inspired broad interest and persistently on-going progress in the exploration of the fascinating optical physics and the advanced OAM-based photonic technology, \cite{shen2019optical,bliokh2015transverse} such as multi-dimensional quantum entanglements,  \cite{erhard2018twisted} OAM-encoded quantum communications, \cite{willner2021orbital,alicia2017free} optical control of microscopic systems, \cite{padgett2011tweezers, Quin2015gauge} and high-resolution imagings. \cite{kozawa2018superresolution}
Following the state-of-the-art advancement in the technology of TL, it is timely crucial to study the light-matter interactions between TL's and opto-electronic materials for the prospective development of TL-based opto-electronic systems. However, the research on the photo-excitation of TL's in solids, especialy the emergent 2D materials, are yet still very limited. \cite{ueno2009coherent,shigematsu2016coherent}  

\begin{figure}[t]
\includegraphics[width=0.95\columnwidth]{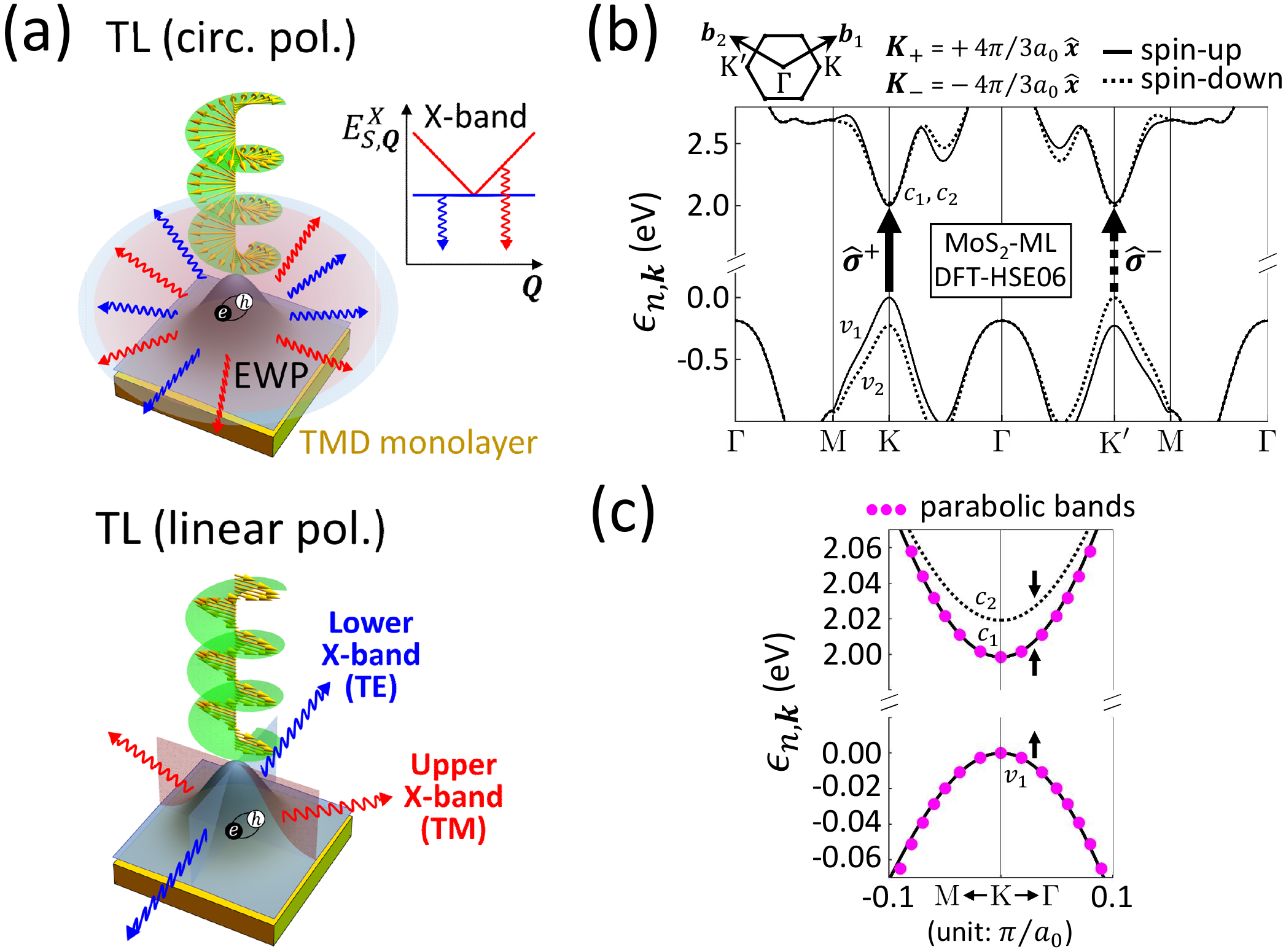}
\caption{(a) Schematics of TMD-ML's under the photo-excitations of twisted lights (TL's) carrying OAM (the spiral wavefront in green color) and SAM, i.e. polarization (the arrow lines in yellow color). Inset: The dipole-allowed bright exciton bands split by {\it e-h} exchange interaction into the linear upper band (red) with longitudinal dipoles and the weakly dispersive lower one (blue) with transverse dipoles. The exciton wave packets (EWP's) photo-generated by the circularly (upper panel) and linearly (lower panel) polarized TL's yield the directional PL's following the completely different angle-dependences. Remarkably, the photo-excitation of {\it linearly} polarized TL sets up an exciton complexer based on TMD-ML, allowing for  resolving and detecting the TM- and TE-luminescences from the upper and lower exciton bands, respectively. (b) The DFT-calculated spin- and valley-characteristic quasi-particle band structure of MoS$_{2}$-ML. (c) The zoom-in band dispersion of the lowest conduction ($\epsilon_{c_1, \vect{k}}$) and topmost valence bands ($\epsilon_{v_1, \vect{k}}$) around the $K$ valley, which can be well fitted by the ideally parabolic bands (magenta dots) with effective mass $m _{c} = 0.39 m _{0}$ and $m _{v} = 0.46 m _{0}$, respectively. For MoS$_2$-ML, the bright A-exciton states involve mainly the spin-like $v_1$ and the $c_1$ bands.}
\label{fig1}
\end{figure}

The first observation of the interaction between TL's and TMD-ML's was reported very recently by Ref. \cite{simbulan2021selective}, with our theoretical contributions. Following the pioneering experiment-theory-joint work, in this paper we for the first time present a comprehensive theoretical investigation of the photo-excited excitons in TMD-ML's by polarized Laguerre Gaussian beams, one of the best known TL's carrying the optical angular momenta, both OAM and SAM (see the schematics shown in Fig.\ref{fig1}(a) for illustration).
We show that the photo-excitations of TL's lead to the formation of excitonic wave packets (EWPs) in TMD-ML's, constituted by the OAM-dependent superposition of the exciton states with different CoM momenta.
Via the couplings between optical angular momenta and the momentum-dependent transition dipoles of valley exciton, TL-induced exciton wave packets in TMD-ML's are shown to yield profound directional photo-luminescences as the manifestation of the intriguing interplay between the multiple degrees of freedom of TL and valley exciton, including both the optical OAM and SAM.  Our studies unveil that the complete exciton-light interaction should be based on the full correspondence between the excitonic and photonic degrees of freedom, including the valley pseudo-spin and center-of-mass motion of exciton, and the corresponding polarization and spatial structure of light. 

This article is organized as follows. Following the introductory section, Section II systematically presents the fundamental theories of valley exciton in 2D materials, the light-matter interactions between 2D excitons and TL's, and the photo-excitation of exciton by TL's. We show how to establish the exciton theory from the fundamental Bethe-Salpeter equation and, for bright exciton states with small CoM momenta, parametrized to be the exciton pseudo-spin model on the first principles base. In the valley-exciton model, we present the theory of exciton-light interaction between the valley exciton states of MoS2$_2$-ML and the Laguerre-Gaussian TL's in the angular spectrum representation.  Taking the time-dependent perturbation theory, we derive the formalisms for the TL-induced formation of exciton wave packets and simulate the resulting angle-resolved photo-luminescences. Sec.III presents and analyzes the calculated results, including the photo-excitation of exciton wave packets by TL's with controlled OAM and SAM, and the OAM- and SAM-encoded shape geometries and angle-resolved photo-luminescence spectra of TL-generated exciton wave packets. Sec. IV concludes this work. 

\section{Theory}

\subsection{Valley-characteristic band structures of TMD-ML's}

The band structure of a TMD-ML is characterized by two distinctive valleys located at the $K$ and $K'$ corners of the first Brillouin zone (BZ), where the conduction and valence bands are separated by the direct band gaps in the visible light regime.
Figure.\ref{fig1}(b) presents the spin-resolved quasi-particle band structure, $\epsilon _{n , \vect{k}}$, of a MoS$_2$ monolayer calculated by using the first principles VASP package \cite{kresse1996efficient} in the density-functional-theory (DFT) with the use of Heyd-Scuseria-Ernzerhof (HSE) functional model. \cite{perdew1996generalized,heyd2003hybrid,heyd2004efficient,krukau2006influence}
Because of the strong spin-orbit couplings (SOCs) in TMD-ML's, the conduction (valence) bands, $\epsilon _{c , \vect{k}}$ ($\epsilon _{v , \vect{k}}$), in the valleys are spin-split by $ \Delta_{c}$ ($\Delta_{v}$) in the scale of few (hundreds of) meV.
The giant spin-splitting in the valence band spectrally separates apart the two classes of exciton states, i.e. the A-exciton (B-exciton) in the low (high) energy regime.
The meV-scaled spin-splitting $\Delta_c$ further lifts the degeneracy of the spin-allowed bright states and the spin-forbidden dark ones in the excitonic fine structure of A-exciton. Throughout this work, we shall focus on the low-lying spin-allowed A-exciton states of MoS$_2$-ML that involve mainly the topmost valence band and lowest spin-like conduction band around the $K$- and $K'$-valleys. For wide-band-gap TMD-ML's, the energy dispersion of the conduction electron states nearby the $K$ ($K'$) valley can be well described by the parabolic bands, $\epsilon_{c , \vect{k}} = E _{g} + \frac{\hbar ^{2} (\vect{k}-\vect{K} _{+}) ^{2}}{2 m _{c}}$ ($\epsilon_{c , \vect{k}} = E _{g} + \frac{\hbar ^{2} (\vect{k}-\vect{K} _{-}) ^{2}}{2 m _{c}}$), where $m _{c}$ is the effective mass of conduction electron, $E _{g}$ is the energy gap, $\vect{K} _{+} = \frac{4 \pi}{3 a _{0}} \hat{\vect{x}}$ ($\vect{K} _{-} = - \frac{4 \pi}{3 a _{0}} \hat{\vect{x}}$) is the position vector of $K$($K ^{\prime}$)-point in the reciprocal space, and $a _{0}=0.3166 $nm is the lattice constant.
Likewise, the energy dispersion of the topmost valence band in the $K$ ($K'$) valley is described by $\epsilon_{v , \vect{k}} = - \frac{\hbar ^{2} (\vect{k} - \vect{K} _{+} ) ^{2}}{2 m _{v}}$ ($\epsilon_{v , \vect{k}} = - \frac{\hbar ^{2} (\vect{k} - \vect{K} _{-}) ^{2}}{2 m _{v}}$), characterized by the effective mass of the valence states, $m _{v}$. From the DFT-calculated band structure of MoS$_{2}$-ML, the effective mass of lowest conduction (topmost valence) band in the $K$-valley is evaluated by using the least square method and fitting to the parabolic bands along the $\Gamma$-$K$ path. Figure.\ref{fig1}(c) shows the DFT-calculated dispersions of the lowest conduction and the topmost valence bands around the $K$-valley of MoS$_2$-ML, well fitted by the parabolic conduction and valence bands with the effective masses, $m_c=0.39 m_0$ and $m_v=0.46 m_0$, respectively, where $m _{0}$ is the electron rest mass.

\subsection{Valley Excitons in TMD-ML's}

\subsubsection{Fundamental exciton theory}
Consider the exciton state with the well-defined center-of-mass (CoM) momentum $\vect{Q}$,
$\left| \Psi _{S , \vect{Q}} ^{X} \right\rangle \equiv \frac{1}{\sqrt{\Omega}} \sum _{v c \vect{k}} \Lambda _{S , \vect{Q}} (v c \vect{k}) \, \crop{c}{c , \vect{k} + \vect{Q}} \, \crop{h}{v , -\vect{k}} \left| GS \right\rangle$, where $S$ is the band index, $ \Omega $ is the total area of the 2D material, $\crop{c}{c , \vect{k }}$($\crop{h}{v , -\vect{k}}$) is the particle operator that creates a conduction electron (valence hole) in the Bloch state $\psi _{c , \vect{k }} (\vect{r})$ ($ \psi _{v , \vect{k}} (\vect{r})$), $ \left| GS \right\rangle $ is the ground state with fully occupied valence bands, and $ \Lambda _{S , \vect{Q}} (v c \vect{k}) $ is the amplitude of the {\it e-h} configuration $\crop{c}{c , \vect{k} + \vect{Q}} \, \crop{h}{v , \vect{-k}} \left| GS \right\rangle$. The energy band structure of an exciton in a TMD-ML is calculated by solving the Bethe-Salpeter equation
\begin{align}\label{eqn:BSE}
\sum _{v ^{\prime} c ^{\prime} \vect{k} ^{\prime}} \left[ \left( \epsilon _{c ^{\prime} , \vect{k} ^{\prime} + \vect{Q}} - \epsilon _{v ^{\prime} , \vect{k} ^{\prime}} \right) \delta_{v, v ^{\prime}} \delta_{c, c ^{\prime}} \delta_{\vect{k}, \vect{k} ^{\prime}} +  U _{\vect{Q}} \! \left( v c \vect{k} , v ^{\prime} c ^{\prime} \vect{k} ^{\prime} \right) \right] \Lambda _{S , \vect{Q}} \! \left( v^{\prime} c^{\prime} \vect{k} ^{\prime} \right) = E_{S , \vect{Q}} ^{X} \, \Lambda_{S , \vect{Q}} \! \left( v c \vect{k} \right)\, ,
\end{align}
where $\epsilon _{c/v, \vect{k}}$ represents the energy of conduction/valence Bloch state $\psi _{c/v , \vect{k}} (\vect{r})$, the kernel of the {\it e-h} Coulomb interaction
$  U _{\vect{Q}} \! \left( v c \vect{k} , v ^{\prime} c ^{\prime} \vect{k} ^{\prime} \right) = - V_{\vect{Q}} ^{d} \! \left( v c \vect{k} , v ^{\prime} c ^{\prime} \vect{k} ^{\prime} \right) + V _{\vect{Q}} ^{x} \! \left( v c \vect{k} , v ^{\prime} c ^{\prime} \vect{k} ^{\prime} \right)$
consists of the screened {\it e-h} direct Coulomb interaction and the {\it e-h} exchange one whose matrix elements are defined by $V _{\vect{Q}} ^{d} \! \left( v c \vect{k} , v ^{\prime} c ^{\prime} \vect{k} ^{\prime} \right) = \, \int d ^{3} \vect{r} _{1} d ^{3} \vect{r} _{2} \, \psi _{c , \vect{k} + \vect{Q}} ^{*} \! \left( \vect{r} _{1} \right) \psi _{v , \vect{k}} \! \left( \vect{r} _{2} \right) W \left( \vect{r} _{1},\vect{r} _{2} \right) \psi _{v ^{\prime} , \vect{k} ^{\prime}} ^{*}  \! \left( \vect{r} _{2} \right) \psi _{c ^{\prime} , \vect{k} ^{\prime} + \vect{Q}}  \! \left( \vect{r} _{1} \right)$ and $V _{\vect{Q}} ^{x} \! \left( v c \vect{k} , v ^{\prime} c ^{\prime} \vect{k} ^{\prime} \right) = \, \int d ^{3} \vect{r} _{1} d ^{3} \vect{r} _{2} \, \psi _{c , \vect{k} + \vect{Q}} ^{*} \! \left( \vect{r} _{1} \right) \psi _{v , \vect{k}}  \! \left( \vect{r} _{1} \right) V \left( \vect{r} _{1} - \vect{r} _{2} \right) \psi _{v ^{\prime} , \vect{k} ^{\prime}} ^{*} \! \left( \vect{r} _{2} \right) \psi _{c ^{\prime} , \vect{k} ^{\prime} + \vect{Q}}  \! \left( \vect{r} _{2} \right)$, respectively. $V \left( \vect{r}_1 - \vect{r}_2 \right)=\frac{e^{2}}{4\pi\varepsilon_{0}|\vect{r}_{1}-\vect{r}_{2}|}$ and $W \left( \vect{r} _{1} , \vect{r} _{2} \right) = \int d ^{3} \vect{r} ^{\prime} \, \varepsilon ^{-1} \left( \vect{r} _{1} , \vect{r} ^{\prime} \right) V \left( \vect{r} ^{\prime} - \vect{r} _{2} \right)$ denotes the bare and screened Coulomb potentials, respectively, where $\varepsilon_{0}$ is the vacuum permittivity, and $\varepsilon ^{-1} \left( \vect{r} _{1} , \vect{r} ^{\prime} \right)$ is the inverse dielectric function.

\subsubsection{Pseudo-spin model for valley exciton}

First, let us disregard the {\it e-h} exchange interaction for exciton and write an exchange-free exciton state of TMD-ML's with the well-specified valley ($\tau=K \text{ or } K ^{\prime}$) character as
\begin{equation}
\left| \Psi _{\tau=K / K ^{\prime}, \vect{Q}} ^{X (0)} \right\rangle = \frac{1}{\sqrt{\Omega }} \sum _{v c \vect{k}} \Lambda _{\tau = K / K ^{\prime} , \vect{Q}}^{(0)} \! \left( v c \vect{k} \right) \op{c} _{c , \vect{k} + \vect{Q}} ^{\dagger} \, \op{h} _{v , - \vect{k}} ^{\dagger} \left| GS \right\rangle \, ,
\label{spin_basis}
\end{equation}
where the superscript $(0)$ is used to indicate the exchange-free nature and the label of exciton state is changed to $S \rightarrow \tau = K / K ^{\prime}$ to specify the valley of exciton.
Note that the exchange-free exciton states are considered to be subjected to no valley-intermixing and to possess the well-specified valley character of exciton.
Taking the long-range approximation for the direct Coulomb interaction \cite{peng2019distinctive} and assuming the constant dielectric function, $\overline{\varepsilon}_b$, Eq.(\ref{eqn:BSE}) can be simplified to the solvable Wannier equation of exciton,
\begin{equation}
\left[ -\frac{\hbar ^{2}}{2 \mu _{X}} \vect{\nabla} _{\!\! \vect{r} \! _{eh}} ^{\, 2} - \frac{e ^{2}}{4 \pi \varepsilon _{0} \bar{\varepsilon} _{b}} \frac{1}{| \vect{r} \! _{eh} |} \right] F _{\tau , \vect{Q}} ^{X (0)}(\vect{r}_{eh}) = \left( E _{\tau , \vect{Q}} ^{X (0)} - E _{g} - \frac{\hbar ^{2} \vect{Q} ^{2}}{2 M _{X}} \right) \, F _{\tau , \vect{Q}} ^{X (0)}(\vect{r}_{eh}) \, ,
\end{equation}
which describes an exciton with well-defined center-of-mass momentum, $\hbar \vect{Q}$, as a hydrogen-atom-like quasiparticle with the reduced mass $\mu\equiv \mu_X =1/(m_c^{-1} + m_v^{-1})$ and the total mass of exciton, $M _{X} = m _{c} + m _{v}$. \cite{fuchs2008efficient,latini2015excitons,berkelbach2015bright} Here, $F _{K / K ^{\prime} , \vect{Q}} ^{X (0)}(\vect{r}_{eh}) = \frac{1}{\Omega} \sum_{\vect{k}} \Lambda _{K / K ^{\prime} , \vect{Q}} ^{(0)} (v c \vect{k} +  \vect{K} _{+ / -} ) \, e ^{i \left( \vect{k} + \frac{\mu _{X}}{m _{c}} \vect{Q} \right) \cdot \vect{r}_{eh}}$
is defined as the envelope wave function of exciton in the relative coordinates, $\vect{r} _{eh} = \vect{r} _{e} - \vect{r} _{h}$.

Within the hydrogen model, the energy dispersion and the momentum-space wave function of the exchange-free lowest $1s$ exciton states are solved as \cite{shibuya1965kepler,parfitt2002two}
\begin{align}\label{X_band_valley}
  E _{\tau , \vect{Q}}^{X (0)} = \left( E _{g} - 4 Ry ^{X} \right) + \frac{\hbar ^{2} \vect{Q} ^{2}}{2 M _{X}} \, ,
\end{align}
and
\begin{align}
\Lambda _{K / K ^{\prime} , \vect{Q}}^{(0)} \! \left( v c \vect{k} \right)
&= \frac{\sqrt{2 \pi} a _{B} ^{X}}{\left[ 1 + \left( \frac{a _{B} ^{X}}{2} \right) ^{2} \left| (\vect{k} - \vect{K} _{+ / -} )+ \frac{\mu _{X}}{m _{c}} \, \vect{Q}  \right| ^{2}  \right] ^{3/2}} \, ,
\label{1s_exciton_wavefunction}
\end{align}
respectively, where $a _{B} ^{X} = \left( 4 \pi \varepsilon _{0} / e ^{2} \right) \left( \hbar ^{2} \bar{\varepsilon} _{b} / \mu _{X} \right)$ is the exciton Bohr radius and $Ry ^{X} \equiv \left( 2 \mu _{X} \right) ^{-1} \left( \hbar / a _{B} ^{X} \right) ^{2}$ is the Rydberg constant.

With the DFT-calculated effective masses, $m _{c} = 0.39 m _{0}$ and $m _{v} = 0.46 m _{0}$, $\mu _{X} = 0.21 m _{0}$ and $M _{X} = 0.85 m_{0}$ are determined. Following the study of Ref.\cite{peng2019distinctive}, we adopt the dielectric constant $\bar{\varepsilon}_{b} =4.6$, yielding the exciton Bohr radius $a _{B} ^{X}= 1.16$ nm, $Ry^{X} = 135$ meV and the binding energy of exciton $E _{b} ^{X (0)} =4 Ry^{X} = 540$ meV.

\begin{figure}[t]
\includegraphics[width=0.9\columnwidth]{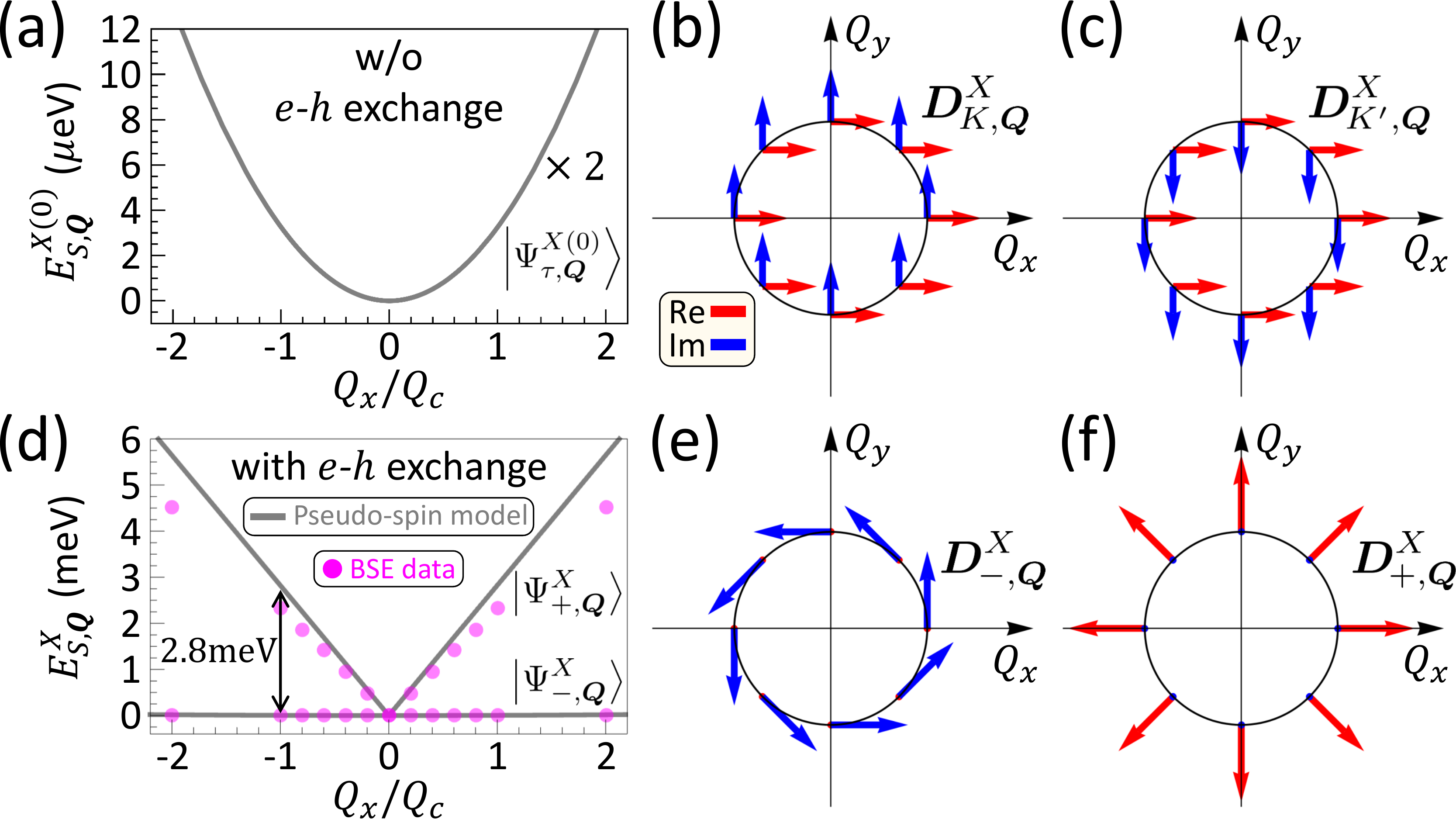}
\caption{(a) The lowest valley-degenerate exciton bands of MoS2$_2$-ML, regardless of {\it e-h} exchange interaction, along the $Q_x$-axis around the light-cone reciprocal range, where $\vect{Q}=(Q_x,Q_y)$ is the wavevector of exciton and the band edge is offset to be zero. (b) The complex transition dipole moments of the $K$-valley exciton states and (c) those of the $K ^{\prime}$-valley exciton states over the $\vect{Q}$-space. The red (blue) arrows denote the real (imaginary) parts of the excitonic dipole moments. In the absence of {\it e-h} exchange interaction, the dipole moments of the $K$- and $K ^{\prime}$-valley exciton states follow the opposite circular polarization, independent of $\vect{Q}$. (d) The valley-split exciton bands by {\it e-h} exchange interaction of MoS2$_2$-ML, calculated by numerically solving the DFT-based BSE (magenta circles) and simulated by the parametrized pseudo-spin model (gray curves). The {\it e-h} exchange interaction splits the degenerate exciton bands into the linear upper band ($S=+$) and the parabolic lower one ($S=-$), leading to the $\vect{Q}$-dependent dipole moments of the exciton states, $\vect{D}_{S,\vect{Q}}^{X}$.  (e) The $\vect{Q}$-dependent transition dipole moments, $\vect{D}_{-,\vect{Q}}^{X}$, of the lower exciton band are shown to be transverse with respect to the $\vect{Q}$ wavevector, while (f) the dipole moments, $\vect{D}_{+,\vect{Q}}^{X}$, of the upper exciton band states are longitudinal.}
\label{fig2}
\end{figure}

Taking the exchange-free exciton doublet as basis, $\left| \Psi _{K, \vect{Q}} ^{X (0)} \right\rangle$ and $\left| \Psi _{K' , \vect{Q}} ^{X (0)} \right\rangle$, the exciton Hamiltonian can be expressed in the form of $2\times 2$ matrix, as given by \cite{yu2014dirac,qiu2015nonanalyticity,peng2019distinctive}
which reads
\begin{equation}
\op{H} _{X} \! \left( \vect{Q} \right) = \left(
 \begin{array}{cc}
  E_{K , \vect{Q}}^{X (0)} + \widetilde{\Delta} _{K , K} \! \left( \vect{Q} \right) & \widetilde{\Delta} _{K , K ^{\prime}} \! \left( \vect{Q} \right) \\
  \widetilde{\Delta} _{K , K ^{\prime}} ^{*} \! \left( \vect{Q} \right) & E _{K ^{\prime}, \vect{Q}} ^{X (0)} + \widetilde{\Delta} _{K ^{\prime} , K ^{\prime}} \! \left( \vect{Q} \right)
 \end{array}
\right) \label{H_X_2x2} ,
\end{equation}
where the matrix elements of \textit{e-h} exchange Coulomb interaction are defined by
\begin{equation}
\widetilde{\Delta} _{\tau , \tau'} \left( \vect{Q} \right) =
\frac{1}{\Omega} \sum _{v c \vect{k}} \sum _{v ^{\prime} c ^{\prime} \vect{k} ^{\prime}} \Lambda_{\tau , \vect{Q}} ^{(0) *} \left( v c \vect{k} \right) V _{\vect{Q}} ^{x} \left( v c \vect{k} , v ^{\prime} c ^{\prime} \vect{k} ^{\prime} \right) \Lambda_{\tau ^{\prime} , \vect{Q}} ^{(0)} \left( v ^{\prime} c ^{\prime} \vect{k} ^{\prime} \right) \, , \label{V_x_SSp}
\end{equation}
where $\tau,\tau' = K \text{ or } K'$.

Considering that bright exciton states lying in the small reciprocal area of light-cone and expanding the periodic part of Bloch function as $u _{n , \vect{k} + \vect{Q}} \left( \vect{r} \right) \approx u _{n , \vect{k}} \left( \vect{r} \right) + \vect{Q} \cdot \vect{\nabla} _{\vect{k}} \, u _{n , \vect{k}} \left( \vect{r} \right)$, \cite{yu2014dirac,qiu2015nonanalyticity,peng2019distinctive} the matrix element of \textit{e-h} exchange Coulomb interaction is approximated to
\begin{equation}\label{ehexch_mtx_elem}
\widetilde{\Delta} _{\tau , \tau ^{\prime}} \! \left( \vect{Q}  \right) \approx \frac{1}{\Omega} \, \frac{1}{2 \varepsilon _{0} \! \left| \vect{Q} \right|} \left( \vect{Q} \cdot \vect{D} _{\tau , \vect{Q}} ^{X *} \right) \left( \vect{Q} \cdot \vect{D} _{\tau ^{\prime} , \vect{Q}} ^{X} \right) ,
\end{equation}
in terms of the transition dipole moment of exciton  $\vect{D} _{\tau , \vect{Q}} ^{X} \equiv \frac{1}{\sqrt{\Omega}} \sum _{v c \vect{k}} \Lambda _{\tau , \vect{Q}}^{(0)} \left( v c \vect{k} \right) \vect{d} _{v , \vect{k} ; c , \vect{k}}$, where $\vect{d} _{v , \vect{k} ; c , \vect{k}} \equiv e \langle \psi _{v , \vect{k}} | \vect{r} | \psi _{c , \vect{k}} \rangle = \frac{e \hbar}{i m _{0} (\epsilon _{v , \vect{k}} - \epsilon _{c , \vect{k}})} \langle \psi _{v , \vect{k}} | \vect{p} | \psi _{c , \vect{k}} \rangle $ is the single-particle transition dipole moment.
For the free electron excitation in the $K$($K'$) valley, it is known that the single-particle transition dipole moment follows the optical $\hat{\vect{\sigma}} ^{+}$($\hat{\vect{\sigma}} ^{-}$)-helicity and written as $\vect{d} _{v , \vect{k} ; c , \vect{k}} = d _{0} \, e ^{-i \phi_{\vect{k} - \vect{K} _{\! +}}} \hat{\vect{\sigma}} ^{+}$ ($\vect{d} _{v , \vect{k} ; c , \vect{k}} = d _{0} ^{\, \ast} \, e ^{i \phi_{\vect{k} - \vect{K} _{\! -}}}\hat{\vect{\sigma}} ^{-}$), where $d _{0}$ is the constant of the average transition dipole moment in the $K$ valley, and $\hat{\vect{\sigma}} ^{\pm} = \frac{1}{\sqrt{2}} (\hat{\vect{x}} \pm i \hat{\vect{y}})$.
With the amplitude $\Lambda _{\tau , \vect{Q}}^{(0)} \left( v c \vect{k} \right)$ of exciton wave function in the $\vect{k}$ space given by Eq.(\ref{1s_exciton_wavefunction}),
we have that $\vect{D} _{K , \vect{Q}} ^{X} =  D _{0} ^{X} \hat{\vect{\sigma}}^{+}$ and $\vect{D}_{K' , \vect{Q}} ^{X} = D _{0} ^{X} \hat{\vect{\sigma}}^{-}$, where $D _{0} ^{X} \equiv \sqrt{\frac{8}{\pi} } \, \frac{\sqrt{\Omega}} {a _{B} ^{X}} \, | d _{0} | $. From the DFT-calculated band structure and the DFT-based BSE calculation (see SI for details), \cite{peng2019distinctive,mostofi2014an,pedersen2001optical} we obtain $a _{B} ^{X} = 1.16$ nm, $| d _{0} | = (0.132 \text{ nm}) |e|$, and $D _{0} ^{X} / \sqrt{\Omega} = 0.181 |e|$.

Accordingly, Eq.(\ref{ehexch_mtx_elem}) is expressed as
\begin{align}
\widetilde{\Delta} _{K , K } \left( \vect{Q}  \right) & = \widetilde{\Delta} _{K ^{\prime} , K ^{\prime}} \left( \vect{Q}  \right) = \gamma \left| \vect{Q} \right| \\
\widetilde{\Delta} _{K , K ^{\prime}} \left( \vect{Q}  \right) & = \widetilde{\Delta} _{K ^{\prime} , K} ^{*} \left( \vect{Q}  \right) = \gamma \left| \vect{Q} \right| e ^{- i 2 \phi _{\vect{Q}}} ,
\end{align}
where $\gamma \equiv \frac{1}{4 \varepsilon _{0}} \left( \frac{D _{0} ^{X}}{\sqrt{\Omega}} \right) ^{2}$ is the strength factor of the {\it e-h} exchange interaction, and $\phi _{\vect{Q}} = \tan ^{-1} (Q _{y} / Q_{x})$.

By diagonalizing Eq.(\ref{H_X_2x2}), the valley-mixed exciton states are solved as
\begin{align}\label{valley_mixed_X_states}
  \left| \Psi _{\pm , \vect{Q}} ^{X} \right\rangle
  = \frac{1}{\sqrt{2}} \left( e ^{-i \phi _{\vect{Q}}} \left| \Psi _{K , \vect{Q}} ^{X (0)} \right\rangle \pm \, e ^{i \phi _{\vect{Q}}} \left| \Psi _{K ^{\prime} , \vect{Q}} ^{X (0)} \right\rangle \right),
\end{align}
with the eigen energies
\begin{align}\label{valley_mixed_X_bands}
E _{\pm , \vect{Q}} ^{X} = E_{\tau , \vect{Q}}^{X(0)} + (1 \pm 1) \, \gamma \left| \vect{Q} \right|\, ,
\end{align}
which are split by the $\vect{Q}$-dependent exchange interaction into the linear upper band $E _{+ , \vect{Q}} ^{X}$ and the parabolic lower band $E _{- , \vect{Q}} ^{X}$, respectively.
The slope of the linear upper band is evaluated as $\gamma\approx 1.47$ eV$\cdot${\AA} from the DFT-calculated $D _{0} ^{X} / \sqrt{\Omega}$, \cite{simbulan2021selective} leading to the exciton band splitting $\sim 2.8$ meV at the light cone edge (see Fig.\ref{fig2}(d)). Following Eq.(\ref{valley_mixed_X_states}), the transition dipole moment of the exciton eigen state with the momentum $\vect{Q}$ in the upper band is given by
\begin{equation}\label{D_upper}
\vect{D}_{+ , \vect{Q}} ^{X} = \frac{1}{\sqrt{2}} \left( e ^{-i \phi _{\vect{Q}}} \vect{D} _{K , \vect{Q}} ^{X} + e ^{i \phi _{\vect{Q}}} \vect{D} _{K' , \vect{Q}} ^{X} \right)= D _{0} ^{X} \! \left( \vect{Q} / |\vect{Q}| \right)
\end{equation}
and that of the exciton state with the same momentum in the lower band is
\begin{equation}\label{D_lower}
\vect{D}_{- , \vect{Q}} ^{X} = \frac{1}{\sqrt{2}} \left( e ^{-i \phi _{\vect{Q}}} \vect{D} _{K , \vect{Q}} ^{X} - e ^{i \phi _{\vect{Q}}} \vect{D} _{K' , \vect{Q}} ^{X} \right) = i D _{0} ^{X} \! \left( \vect{Q} _{\perp} / |\vect{Q}| \right)\, ,
\end{equation}
where $\vect{Q}_{\perp}\equiv \left| \vect{Q} \right| \left( - \sin \phi _{\vect{Q}} \, \hat{\vect{x}} + \cos \phi _{\vect{Q}} \, \hat{\vect{y}} \right)$ is perpendicular to $\vect{Q}$. One notes that the exciton dipole moment of upper (lower) exciton band is longitudinal (transverse) with respect to the exciton wavevector, i.e. $\vect{D} _{+, \vect{Q}} ^{X} \parallel \vect{Q}$  ($\vect{D} _{- , \vect{Q}} ^{X} \perp \vect{Q}$).

Figure.\ref{fig2}(d) presents the calculated valley-mixed bright exciton bands spit by the {\it e-h} exchange interaction in the $\vect{Q}$-space covering the light-cone area by using Eq.(\ref{valley_mixed_X_bands}) with the parameter $\gamma = 1.47$ eV$\cdot\text{\AA}$, in agreement with the BSE-calculated band structures. \cite{simbulan2021selective} Correspondingly, the transition dipole moments of the valley-mixed exciton states with finite $\vect{Q}$ in the lower and upper bands are plotted in Fig.\ref{fig2}(e) and (f), respectively. Figure.\ref{fig2}(e) and (f) show that the dipole moments of the valley-mixed excitons are $\vect{Q}$-dependent, where $\vect{D} _{+ / - , \vect{Q}} ^{X}$ of the upper longitudinal/lower transverse band is always pointing to the direction parallel/normal to $\vect{Q}$.
For comparison, Figure.\ref{fig2}(a)-(c) show the calculated valley-degenerate parabolic exciton bands and the exciton dipole moments in circular polarization of bright valley exciton without the consideration of {\it e-h} exchange interaction.

\subsection{Twisted light: Laguerre-Gaussian beams}

Throughout this work, we consider the twisted light in the Laguerre-Gaussian (LG) modes characterized with  the quatized orbital angular momentum (OAM) $ \ell=0,\pm 1,\pm 2,...$ and radial mode index $p=0,1,2,...$ that are normally incident on TMD-ML's. \cite{allen1992orbital}
Experimentally, LG beams with the well-controlled OAMs can be realized by using the technology of spatial light modulator (SLM), and have been employed to photo-excite atoms, \cite{inoue2006entanglement} molecules, \cite{araoka2005interactions,loffler2011circular} and free electrons in crystals, \cite{clayburn2013search} but not extensively yet excitons in solids. \cite{ueno2009coherent,shigematsu2016coherent,simbulan2021selective}
Taking the paraxial approximation (that is valid for not too large $\ell$), the vector potential of the LG beam in Lorenz gauge under the long Rayleigh-length condition \cite{romero2002quantum} is given by
\begin{align}\label{LG_beam}
    \vect{A}_{\vect{q_0}} ^{\hat{\vect{\varepsilon}} , \ell , p} (\vect{r})
    = \hat{\vect{\varepsilon}} \, A _{0} ^{\text{LG}} \, u _{\ell p} (\vect{r} _{\parallel}) \, e ^{i q _{0} z} ,
\end{align}
where $\vect{q}_{0}= q _{0} \hat{\vect{z}}$ is the wavevector specified to be along the direction of light propagation, $\hat{\vect{\varepsilon}}$ denotes the unit vector of polarization perpendicular to $\vect{q} _{0}$, and $A _{0} ^{\text{LG}}$ is the constant amplitude. The spatial distribution of the LG beam with the OAM $\ell$ in transverse direction $\vect{r} _{\parallel} = x \, \hat{\vect{x}} + y \, \hat{\vect{y}}$ is expressed as
\begin{align}\label{amplitude}
  u _{\ell p} (\vect{r} _{\parallel}) = f _{\ell p} (\rho) \, e ^{i \ell \phi} ,
\end{align}
where $f _{\ell p} (\rho) = \sqrt{p ! 2 / ( |\ell| + p ) ! \pi} ~ e ^{- \rho ^{2} / W _{0} ^{2}} ~ ( \sqrt{2} \rho / W _{0} ) ^{| \ell |} ~ L _{p} ^{| \ell |} ( 2 \rho ^{2} / W _{0} ^{2} )$ is the radial function for the vector potential in terms of the generalized Laguerre polynomial $L _{p} ^{| \ell |} (x)$ and parametrized by the beam waist $W _{0}$, $\rho = \sqrt{x ^{2} + y ^{2}}$, and $\phi = \tan ^{-1} (y/x)$. The phase factor $e ^{i \ell \phi}$ for finite OAM, $\ell \neq 0$, is featured by a singularity at the origin where $\phi$ is ill-defined. In the angular spectrum representation, the LG beam in Eq.(\ref{LG_beam})  can be further decomposed in the 2D Fourier transform \cite{sherman1982introduction} as
\begin{align}\label{LG_beam_ASR}
  \vect{A} _{\vect{q} _{0}} ^{\hat{\vect{\varepsilon}} , \ell , p} (\vect{r})
  = \hat{\vect{\varepsilon}} \sum _{\vect{q} _{\parallel}} \mathcal{A} _{\ell p} (\vect{q} _{\parallel}) e ^{i \vect{q} \cdot \vect{r}} ,
\end{align}
where $\vect{q} _{\parallel} = q _{x} \, \hat{\vect{x}} + q _{y} \, \hat{\vect{y}}$ and $\vect{q} = \vect{q _{\parallel}} + \vect{q}_{0} $.
As derived by Ref.\cite{simbulan2021selective} the $\vect{q} _{\parallel}$-space amplitude function of LG beam is given by
\begin{align}\label{amplitude_ASR}
  \mathcal{A} _{\ell p} (\vect{q} _{\parallel}) = \tilde{F} _{\ell p} (q _{\parallel}) \, e ^{i \ell \phi _{\vect{q} _{\parallel}}} \, ,
\end{align}
in terms of the complex-valued radial function
\begin{equation}
\tilde{F} _{\ell p} (q _{\parallel}) = e ^{-i \ell \pi / 2} \, e^ {i \eta _{\ell} \pi} \, F _{\ell p} (q _{\parallel})
\end{equation}
and the phase factor $e ^{i \ell \phi _{\vect{q} _{\parallel}}}$, where $q _{\parallel} = \sqrt{q _{x} ^{2} + q _{y} ^{2}}$, $\phi _{\vect{q} _{\parallel}} = \tan ^{-1} (q _{y} / q_{x})$, $\eta _{\ell} \equiv| \ell | \left[ 1 - \Theta (\ell) \right]$, $\Theta (\ell)$ is the Heaviside function, and the real-valued radial function
\begin{equation}
F _{\ell p} (q _{\parallel}) = (2 \pi A _{0} ^{\text{LG}} / \Omega) \, \mathcal{H} _{| \ell |} \! \left[ f _{\ell p} (\rho) \right]
\label{real_valued_radial_function_of_LG_beam}
\end{equation}
is obtained by means of the Hankel transform $\mathcal{H} _{| \ell |} \! \left[ f _{\ell p} (\rho) \right] = \int _{0} ^{\infty} d \rho \, \rho \, f _{\ell p} (\rho) J _{| \ell |} (q _{\parallel} \rho)$ of order $| \ell |$ with $J _{| \ell |} (q _{\parallel} \rho)$ being the Bessel function of the first kind of order $| \ell |$. \cite{arfken2012mathematical}

\begin{figure}[t]
\includegraphics[width=0.9\columnwidth]{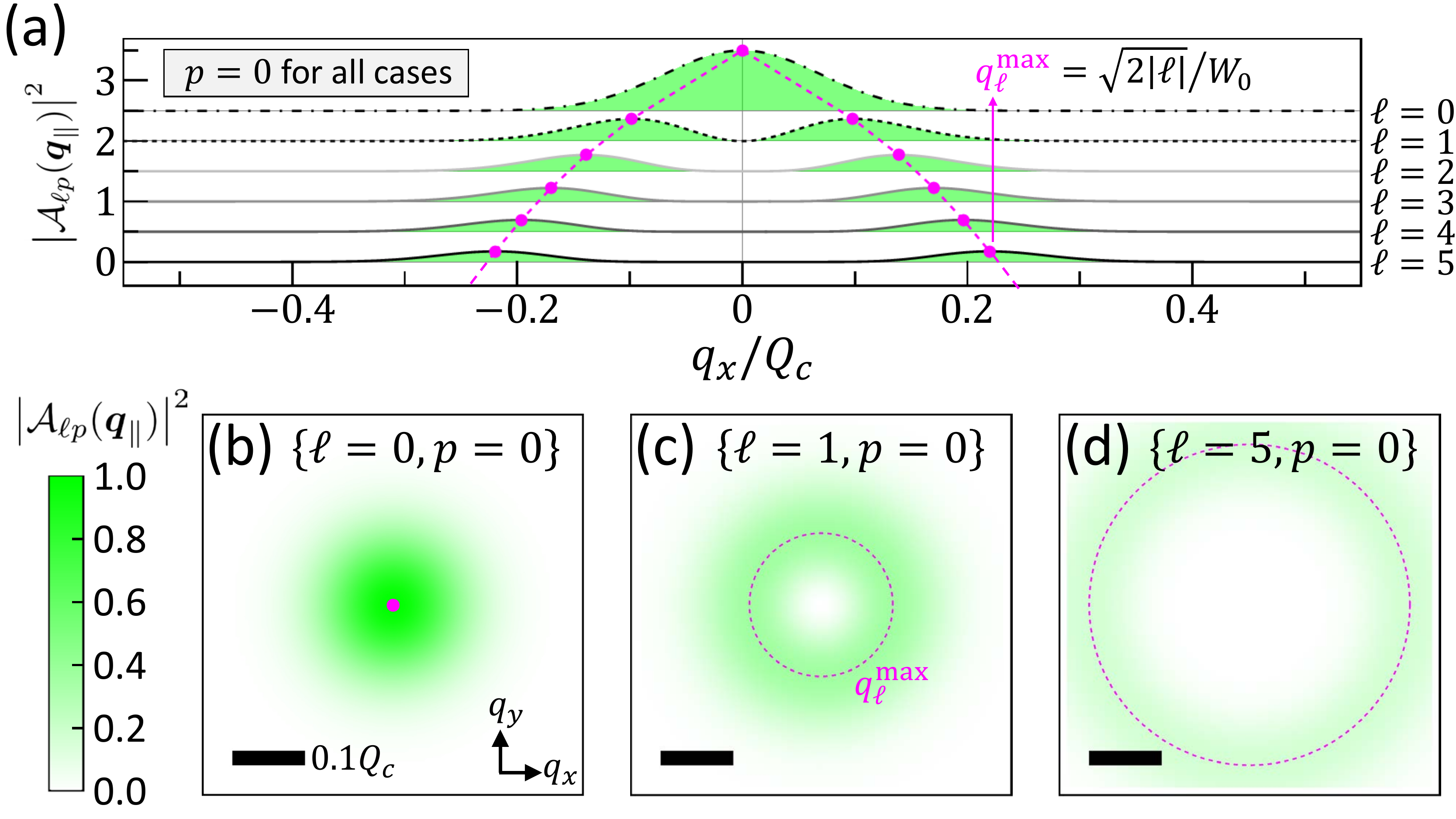}
\caption{(a) The square of the magnitude of the amplitude functions, $\big| \mathcal{A} _{\ell p} (\vect{q} _{\parallel}) \big|^{2}$, of the TL's in the LG modes with the OAM's $\ell=0, 1,..,5$ and $p$ fixed to $0$, plotted along the $q _{x}$-axis. For clarity, the profiles of $\big| \mathcal{A} _{\ell p} (\vect{q} _{\parallel}) \big|^{2}$ for the LG beams of different OAM's are offset by the $\ell$ values. The colored contour plots of the $\big| \mathcal{A} _{\ell p} (\vect{q} _{\parallel}) \big| ^{2}$ over the $\vect{\vect{q}_{\parallel}}$-space centred in the light cone for the LG beams with (b) $\ell = 0$ and $p = 0$, (c) $\ell = 1$ and $p = 0$, and (d) $\ell = 5$ and $p = 0$. In (b)-(d), the black scale bar on the lower-left represents the length of $0.1 Q _{c}$ for reference. All the contour plots of $\big| \mathcal{A} _{\ell p} (\vect{q} _{\parallel}) \big| ^{2}$ in (b)-(d) follow the same color-map on the most left-handed side. The magenta dots and dashed contour lines label the wavevector at which the amplitude function is the maximum for a specific $\ell$-mode of TL.}
\label{fig3}
\end{figure}

\begin{figure}[t]
\includegraphics[width=0.9\columnwidth]{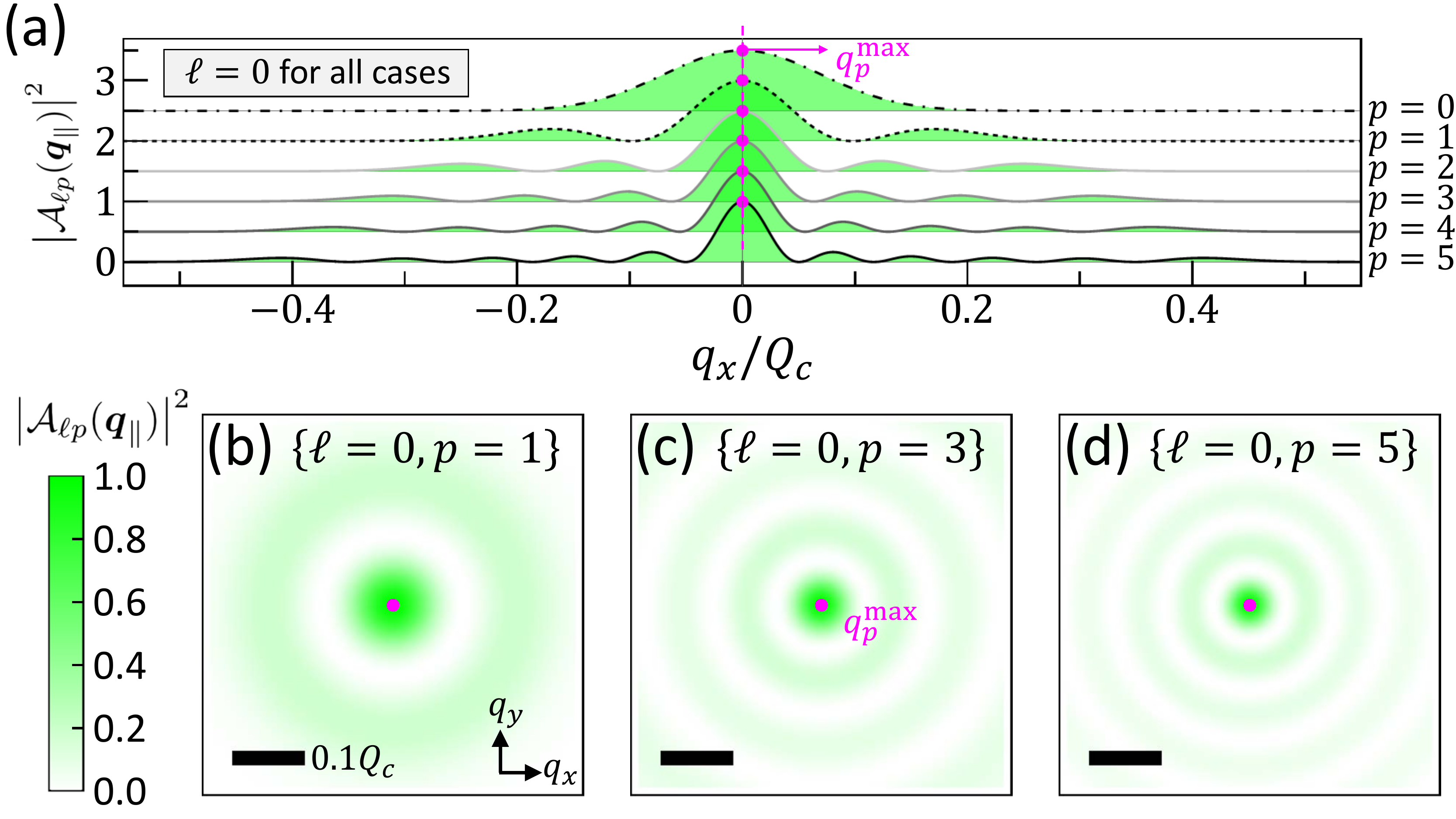}
\caption{(a) The square of the magnitude of the amplitude functions, $\big| \mathcal{A} _{\ell p} (\vect{q} _{\parallel}) \big|^{2}$, of the TL's in the LG modes with the vanishing OAM $\ell=0$ and the various radial indices $p=0, 1,..5$. The colored contour plots of the $\big| \mathcal{A} _{\ell p} (\vect{q} _{\parallel}) \big| ^{2}$ over the $\vect{\vect{q}_{\parallel}}$-space centred in the light cone for the LG beams with (b) $\ell = 0$ and $p = 1$, (c) $\ell = 0$ and $p = 3$, and (d) $\ell = 0$ and $p = 5$. One notes that the amplitude functions of the TL's in the LG modes with non-zero $p$ are featured with the rippling patterns with the function maximum located at the $\vect{q} _{\parallel} = \vect{0}$ point. }
\label{fig4}
\end{figure}

Taking the form of Eq.(\ref{LG_beam_ASR}), a TL can be viewed as a superposition of a large number of the plane-waves with distinct wavevectors $\vect{q} = (\vect{q}_{\parallel}, q_0)$, each of which propagates with the amplitude of Eq.(\ref{amplitude_ASR}) in the slightly different directions inclined from that of $\vect{q} _{0}$ with the angles, $\theta = \tan ^{-1} \left( | \vect{q} _{\parallel} | / | \vect{q} _{0} | \right)$, depending on  $\vect{q} _{\parallel}$ ($\theta \ll 1$ in the paraxial approximation).
As shown later, the amplitude of Eq.(\ref{amplitude_ASR}), $\mathcal{A} _{\ell p} (\vect{q} _{\parallel}) $, determines the magnitude of optical matrix element with which the finite-momentum exciton state with $\vect{Q}=\vect{q} _{\parallel} \neq \vect{0}$ can be photo-excited by TL.
In the limit of small $\theta$, it is shown by Ref.\cite{peshkov2017photoexcitation} that the formalism of LG beam formulated by Eq.(\ref{LG_beam}) in Lorenz gauge can be approximated by that of the same beam in the Coulomb gauge. Under the condition, one can incorporate the formalism of Eq.(\ref{LG_beam_ASR}) for the TL in the paraxial approximation into the light-matter interaction theory presented in the next section which is set up on the base of Coulomb gauge.\cite{Quin2015gauge}

Figure.\ref{fig3}(a) shows the magnitude of the $\vect{q} _{\parallel}$-space amplitude function, $\big| \mathcal{A} _{\ell p = 0} (\vect{q} _{\parallel}) \big| ^{2}$, of the TL's described by Eq.(\ref{amplitude_ASR}) with $\ell=0,1,2...5$ and $W _{0} = 1.5$ $\mu$\text{m}, distributed over the $q _{x}$-axis.
In the connection to the light-exciton interaction analyzed later, Figure.\ref{fig3} plots $\big| \mathcal{A} _{\ell p} (\vect{q} _{\parallel}) \big| ^{2}$ as a function of $\vect{q} _{\parallel}/Q_c$, normalized by the light-cone radius $Q _{c} = (1.92 \, \text{eV}) / \hbar c$, where $1.92 \, \text{eV}$ is the measured energy of A-exciton peak in the absorption spectra of MoS$_{2}$-ML. \cite{mak2013tightly,qiu2015optical}

From Fig.\ref{fig3}, we see that the distribution of $\big| \mathcal{A} _{\ell=0 p=0} (\vect{q} _{\parallel}) \big| ^{2}$ in the $\vect{q} _{\parallel}$-space is centralized around the origin of $\vect{q} _{\parallel} = \vect{0}$, whereas those of $\big| \mathcal{A} _{\ell\neq 0 p=0} (\vect{q} _{\parallel}) \big| ^{2}$ in the $\vect{q} _{\parallel}$-space for the TL's with finite $\ell=1,2,...$ are featured by donut-like distribution where the magnitudes of $\big| \mathcal{A} _{\ell\neq 0 p = 0} (\vect{q} _{\parallel}) \big| ^{2}$ vanish at the origin. This indicates that the TL's with finite $\ell$ are likely to excite the exciton states with finite $\vect{Q}$ as recently reported by Ref.\cite{simbulan2021selective}. One also notes that the wavevector, $q_{\ell}^{\text{max}}$, where the $\big| \mathcal{A} _{\ell \neq 0 p=0} (\vect{q} _{\parallel}) \big| ^{2}$ is maximum increases with increasing the OAM of TL, $\ell \hbar$. In fact, it can be analytically shown that $q _{\ell}^{\text{max}} = \sqrt{2 | \ell |} / W _{0}$ is proportional to $\sqrt{ | \ell |}$ for $p = 0$.

Figure.\ref{fig4}(a) shows the magnitude of the vector potentials as a function of $\vect{q}_{\parallel}$ for the LG beams with $\ell = 0$ and $p =1,2,...\ge 0$, presenting the oscillations featured by $p$-nodes and $(p + 1)$-maxima in the radial direction. Unlike Fig.\ref{fig3}(a), the $q_{\ell=0, p}^{\text{max}}$   where $\big| \mathcal{A} _{\ell = 0 p} (\vect{q} _{\parallel}) \big| ^{2}$ are maximum in Fig.\ref{fig4}(a) are fixed to be $q_{p}^{\text{max}}=0$ for the different $p$ modes of LG TL with $\ell=0$. Figure.\ref{fig4}(b)-(d) show the function, $\big| \mathcal{A} _{\ell = 0 p} (\vect{q} _{\parallel}) \big| ^{2}$, in the $\vect{q} _{\parallel}$-plane. As compared with Fig.\ref{fig3}(b)-(d) for the LG beams with $p=0$, $\big| \mathcal{A} _{\ell = 0 p} (\vect{q} _{\parallel}) \big| ^{2}$ in Fig.\ref{fig4}(b)-(d) exhibit the isotropic and multi-ring structures ($p$ is the number of rings). Besides, one notes that the functions $\big| \mathcal{A} _{\ell = 0 p} (\vect{q} _{\parallel}) \big| ^{2}$ at the origin point where $\vect{q} _{\parallel} = \vect{0}$ is not the nodal point but actually the global maximum of the functions.

\subsection{Exciton-light interaction}

In this section, we present the formalisms for the exciton-light interaction between 2D valley exciton and LG TL formulated in the angle spectrum representation. 
Taking the rotating wave approximation, one writes the light-matter interaction induced by a weak TL as $H _{I}\left( \vect{r} , t \right) \approx \tilde{H} _{I} \left( \vect{r} \right) \, e ^{-i \omega t}$, where
\begin{equation}
  \tilde{H} _{I}\left( \vect{r} \right) =  \frac{\left| e \right|}{2 m _{0}} \vect{A} ( \vect{r} ) \cdot \vect{p}  \, .
  \label{light_matter_interaction_H_RWA}
\end{equation}

In the time-dependent perturbation theory, the time-dependent exciton state under a weak photo-excitation can be expressed as
\begin{equation}
|\Psi^X (t) \rangle \approx |GS\rangle + \sum_{S} \sum _{\vect{Q}} \tilde{c}_{S , \vect{Q}} ^{(1)} (t) e^{-i E _{S, \vect{Q}} ^{X} t/\hbar} \, |\Psi _{S , \vect{Q}} ^{X}\rangle \, , \label{coherent_state}
\end{equation}
where $\tilde{c}_{S , \vect{Q}} ^{(1)} (t) $ are the TL-induced time-dependent coefficients of the photo-generated exciton states, $|\Psi _{S , \vect{Q}} ^{X}\rangle$. In the first-order perturbation theory, one solves
\begin{equation}
\tilde{c}_{S , \vect{Q}} ^{(1)} (t) = \tilde{\gamma}_{S , \vect{Q}}(t) \left( \frac{\tilde{M} _{S , \vect{Q}}}{\hbar \omega _{S , \vect{Q}}} \right) \, ,
\end{equation}
where $\tilde{\gamma}_{S , \vect{Q}}(t)= - 2 i e ^{i (\omega _{S , \vect{Q}} - \omega) t / 2} \, \left( \frac{\sin \left( \left( \omega _{S , \vect{Q}} -\omega \right) t / 2 \right)}{ 1 - \left( \omega / \omega _{S , \vect{Q}} \right) +i 0^{+}} \right) $ absorbs the all time-dependence of the coefficient,
\begin{equation}
\tilde{M} _{S , \vect{Q}} =  \frac{1}{\sqrt{\Omega}} \sum _{v c \vect{k}} \Lambda _{S , \vect{Q}} ^{*} (v c \vect{k}) \langle \psi _{c , \vect{k} + \vect{Q}} | \tilde{H} _{I} (\vect{r}) | \psi _{v , \vect{k}} \rangle
\end{equation}
is the optical matrix element that measures the optical activity of the exciton state $\left| \Psi _{S , \vect{Q}} ^{X} \right\rangle$ with respect to the incident light and $\omega _{S , \vect{Q}} = E _{S , \vect{Q}} ^{X} / \hbar$ is the natural frequency of the exciton state $\big| \Psi _{S , \vect{Q}} ^{X} \big\rangle$. In the long-time limit, the time-dependent part approaches the delta function $\big| \tilde{\gamma}_{S , \vect{Q}}(t) \big| ^{2} \rightarrow 2 \pi t \hbar \, \omega _{S , \vect{Q}} ^{2} \, \delta ( \hbar \omega _{S , \vect{Q}} - \hbar \omega )$, leading to the Fermi's golden rule.

For normally incident non-structured lights, the vector potentials are generally given by $\vect{A} _{\vect{q}_0} ^{\hat{\vect{\varepsilon}}} (\vect{r}) = \hat{\vect{\varepsilon}} A _{0} \, e ^{i \vect{q}_0 \cdot \vect{r}}$, specified by a well-defined wavevector $\vect{q}_0$. Taking the vector potential of plane-wave light for Eq.(\ref{light_matter_interaction_H_RWA}), the optical matrix element of the exciton state $| \Psi _{S , \vect{Q}} ^{X} \rangle$ in the electric dipole approximation is derived as $\tilde{M} _{S , \vect{Q}} ^{\hat{\vect{\varepsilon}}, \vect{q} _{0}} \approx \delta _{\vect{Q} , \vect{0}} \left( E _{g} / 2 i \hbar \right) A_{0} \left( \hat{\vect{\varepsilon}} \cdot \vect{D} _{S , \vect{Q}} ^{X *} \right)$
where the Kronecker delta, $\delta _{\vect{Q} , \vect{0}}$, ensuring the conservation of momentum allows only the exciton state with vanishing $\vect{Q}=\vect{0}$ coupled to the normal incident light with $\vect{q}_{\parallel}=\vect{0}$.

By contrast, a TL is composed of infinite number of finite-momentum plane-waves as presented by Eq.(\ref{LG_beam_ASR}). Thus, the optical matrix element of the exciton state $| \Psi _{S , \vect{Q}} ^{X} \rangle$ in TMD-ML's under the electric dipole approximation is given by,
\begin{equation}\label{optical_matrix_element_LG_beam}
\tilde{M} _{S , \vect{Q}} ^{\hat{\vect{\varepsilon}}, \ell,p}
\approx \left( \frac{E _{g}}{2 i \hbar}  \right) \mathcal{A} _{\ell p} (\vect{Q}) \big( \hat{\vect{\varepsilon}} \cdot \vect{D} _{S , \vect{Q}} ^{X *} \big)\, .
\end{equation}
The $\vect{Q}$-dependent optical matrix element of Eq.(\ref{optical_matrix_element_LG_beam}) is determined by the product of the OAM-determined spectrum distribution function of LG beam, $\mathcal{A} _{\ell p} \left( \vect{Q} \right)$, over the extended $\vect{Q}$-space and the projection of exciton transition dipole moment on the polarization direction, $\left( \hat{\vect{\varepsilon}} \cdot \vect{D} _{S , \vect{Q}} ^{X *} \right)$, i.e. the coupling between the optical SAM, $\hat{\vect{\varepsilon}}$, of TL and the valley-mixed transition dipole of exciton by the intrinsic {\it e-h} exchange interaction. The magnitude of the optical matrix element evaluated by Eq.(\ref{optical_matrix_element_LG_beam}) reflects the coupling strength for the exciton-light interaction between a valley exciton in the state $(S,\vect{Q})$ and a $\hat{\vect{\varepsilon}}$-polarized TL in the  $(\ell,p)$ mode, manifesting the intriguing couplings between the multiple degrees of freedom possessed by the optical and excitonic subsystems.
With the  $\vect{Q}$-dependent non-zero optical matrix elements given by Eq.(\ref{optical_matrix_element_LG_beam}), a LG beam enables the simultaneous photo-generation of numerously distinct finite-moment exciton states, $\left| \Psi _{S , \vect{Q}} ^{X} \right\rangle$, forming a wave packet of exciton.

\subsection{TL's induced exciton wave packets}

To characterize the spatial localization of an exciton wave packet, here we introduce
the envelope function of wave packet in the center-of-mass coordinates $\vect{R}_{\, c}$ defined by
\begin{align}\label{wave_packet}
\tilde{F}_S (\vect{R} _{c},t) = \frac{1}{\sqrt{\Omega}}  \sum _{\vect{Q}} \tilde{c}_{S , \vect{Q}}^{(1)}(t) \, e ^{i\vect{Q} \cdot  \vect{R} _{c}}\, ,
\end{align}
which extracts the $\vect{R} _{c}$-coordinate envelope profile from the exciton wave function given by Eq.\ref{coherent_state}.
Neglecting the slight phase interference between the time-dependent parts ($\propto e^{-i \omega_{S , \vect{Q}} t}$) of the different exciton states and considering $\omega_{S, \vect{Q}}\approx \omega_{\vect{0}}$ to be approximately constant, the amplitude coefficients of the exciton states is written as
$\tilde{c}_{S , \vect{Q}}^{(1)}(t) \approx \tilde{\gamma}_{\vect{0}} (t) \Big( \frac{\tilde{M} _{S , \vect{Q}} ^{\hat{\vect{\varepsilon}} , \ell , p}}{\hbar \omega_{\vect{0}}} \Big) $, where $\tilde{\gamma}_{S , \vect{Q}} (t) \approx \tilde{\gamma}_{S, \vect{0}} (t) \equiv \tilde{\gamma}_{\vect{0}}(t)$.
Thus, the envelope function of exciton wave packet for the $S$ band excited by the $\hat{\vect{\varepsilon}}$-polarized TL in the $(\ell,p)$-mode can be written as $\tilde{F}_S^{\hat{\vect{\varepsilon}}, \ell, p} ( \vect{R} _{c} , t ) \approx \tilde{\gamma}_{\vect{0}}(t) \tilde{f}_S^{\hat{\vect{\varepsilon}}, \ell, p} (\vect{R} _{c})$ in a separable form, where $\tilde{\gamma}_{\vect{0}}(t)$ is the factor absorbing the all time-dependences and
\begin{equation}\label{TID_wave_packet}
\tilde{f}_S^{\hat{\vect{\varepsilon}}, \ell, p} (\vect{R} _{c}) =  \frac{1}{\sqrt{\Omega}} \sum _{\vect{Q}} \tilde{g}_{S}^{\hat{\vect{\varepsilon}}, \ell, p}(\vect{Q}) \, e ^{i\vect{Q} \cdot  \vect{R} _{c}} \, ,
\end{equation}
is defined to describe the static part of the envelope function,
where $\tilde{g}_{S}^{\hat{\vect{\varepsilon}}, \ell, p} (\vect{Q}) \equiv  \left( {\tilde{M}_{S , \vect{Q}}^{\hat{\vect{\varepsilon}}, \ell, p} }/{\hbar \omega_{\vect{0}}} \right)$ is proportional to the optical matrix element and acts as the coefficient of the Fourier transform component of the envelope function, $\tilde{f} _{S} ^{\hat{\vect{\varepsilon}} , \ell , p} (\vect{R} _{c})$, of the TL-induced exciton wave packet in the center-of-mass coordinate $\vect{R} _{c}$-space.
 Thus, the contour pattern of $\tilde{M} _{S , \vect{Q}}^{\hat{\vect{\varepsilon}}, \ell, p} $ over the $\vect{Q}$-space for the exciton wave packet excited by $\hat{\vect{\varepsilon}}$-polarized TL in the $(\ell, p)$ mode directly maps out the finite-momentum exciton components therein, from which one can further infer the shape of the wave packet in the real space described by $\tilde{f} _{S} ^{\hat{\vect{\varepsilon}} , \ell ,p} (\vect{R} _{c})$. In other words, a TL-induced exciton wave packet can be re-shaped by switching the $(\ell , p)$ mode of the LG beam or directly changing the light polarization. In fact, engineering or guiding the center-of-mass motion of an exciton is not a trivial task because of charge neutrality of exciton.\cite{onga2017exciton,yang2021waveguiding,fedichkin2016room} Here, we find that the application of TL provides an alternative possible route to forming and shaping the wave packet of exciton.

In the remainder part of this work, we will investigate how to tailor the geometric shape of the exciton wave packet in TMD-ML's using polarized TL's and analyze the $\ell$- and $p$-dependent angle-resolved optical spectra of the TL-tailored exciton wave packets.

\section{Results and discussions}

\begin{figure}[t]
\includegraphics[width=0.9\columnwidth]{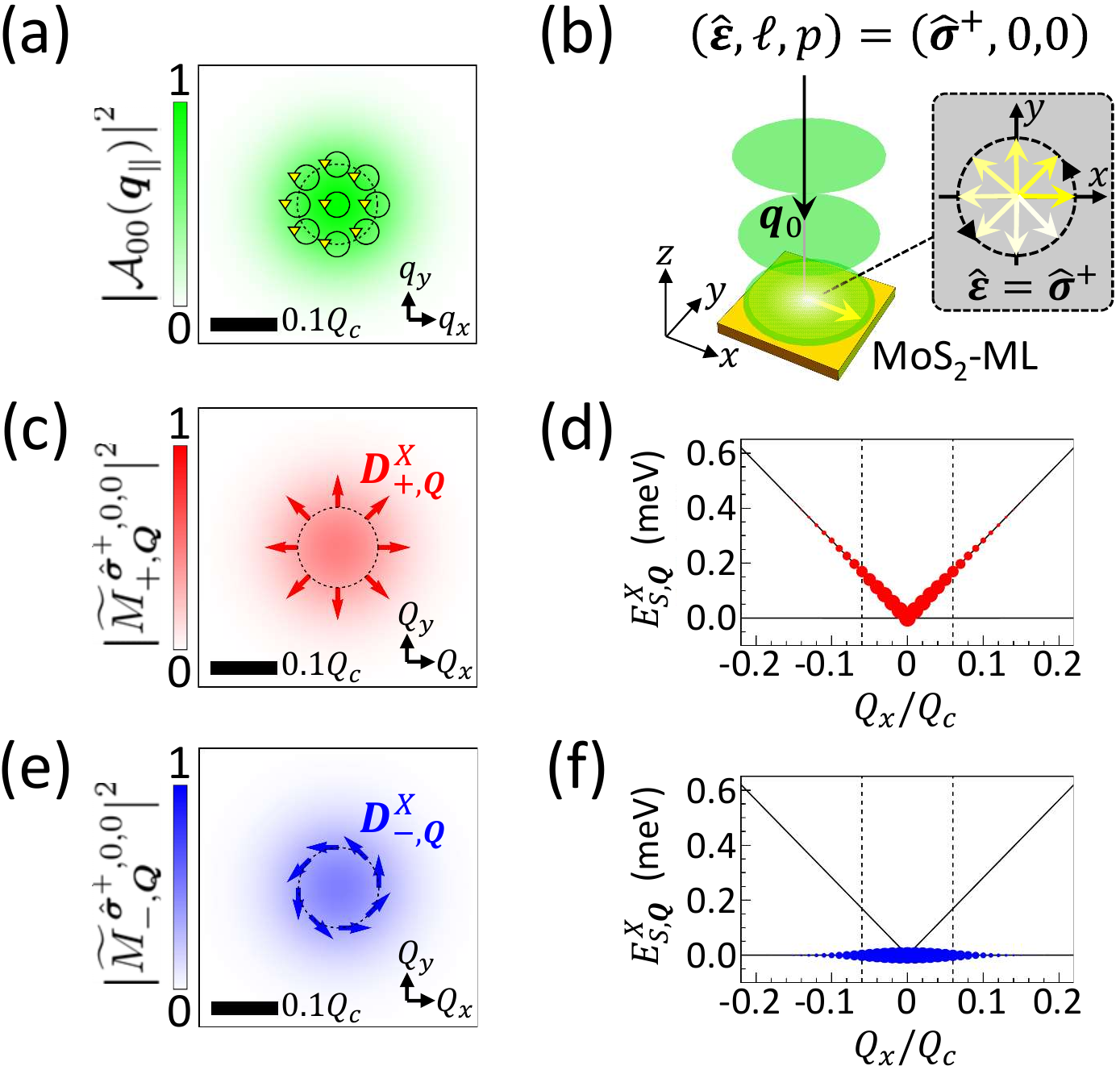}
\caption{(a) The square of the magnitude of the amplitude function, $\big| \mathcal{A} _{\ell p} (\vect{q} _{\parallel}) \big| ^{2}$, of normally incident circularly polarized LG beam in the fundamental mode with $\vect{q} _{0} = q _{0} \hat{\vect{z}}$ and $( \hat{\vect{\varepsilon}} , \ell , p ) = ( \hat{\vect{\sigma}} ^{+} , 0 , 0 )$, where the spiral arrow lines schematically indicate the circular polarization of TL. (b) Schematics of a MoS$_{2}$ monolayer under the photo-excitation of the circularly polarized TL. (c) The square of the magnitude of the optical matrix elements, $\left| \tilde{M} _{S = + , \vect{Q}} ^{\hat{\vect{\varepsilon}} = \hat{\vect{\sigma}} ^{+} , \ell = 0 , p = 0} \right| ^{2}$, of the valley-mixed exciton states in the upper band, where the red arrow lines depict the longitudinal dipole moments, $\vect{D} _{+,\vect{Q}} ^X \parallel \vect{Q}$, of the TL-excited exciton states with $Q=0.06 Q _{c}$ (indicated also by the vertical dashed lines in (d) and (f)). (d) The valley-split exciton band structure and the exciton states photo-generated by the fundamental LG TL in the upper band, marked by  the red filled circles whose sizes reflect the magnitudes of $\left| \tilde{M} _{+ , \vect{Q}} ^{\hat{\vect{\sigma}} ^{+} , 0 , 0} \right| ^{2}$ of the states. (e) and (f) present the results as (c) and (d), but for the exciton states in the lower band. Note the photo-excited exciton states in the lower band carrying the transverse dipole moments, $\vect{D} _{-,\vect{Q}} ^X \perp\vect{Q}$.}
\label{fig5}
\end{figure}

\subsection{Formation of exciton wave packets by TL's}

\subsubsection{Optical matrix elements of excitons and the reciprocal amplitude functions of LG TL's}

First, we consider the  exciton states photo-generated by normally incident polarized LG beams in the fundamental mode $(\ell,p)=(0,0)$.
Figure.\ref{fig5}(c) and (e) show the square of magnitude of the optical matrix elements, $\Big| \tilde{M} _{S=\pm , \vect{Q}} ^{\hat{\vect{\sigma}} ^{+} , 0 , 0} \Big|^{2}$, for the $\vect{Q}$-exciton components in the upper and lower exciton band, respectively, excited by a circularly $\hat{\vect{\sigma}} ^{+}$-polarized fundamental $(0,0)$-LG beam.

From Eq.(\ref{optical_matrix_element_LG_beam}), one realizes that the complex optical matrix elements, $\tilde{M} _{S , \vect{Q}} ^{\hat{\vect{\varepsilon}} , \ell , p} $ for the $\vect{Q}$-exciton components in a TL-generated exciton wave packet are determined by the kind of mode of the exciting LG beam as well as the polarization $\hat{\vect{\varepsilon}}$. This indicates that the superposition of the $\vect{Q}$-exciton components in a exciton wave packet can be specified by the selected polarization, $\hat{\vect{\varepsilon}}$, of incident TL. 
Thus, because the magnitude of the dipole-coupling term $\left| \hat{\vect{\sigma}} ^{+} \cdot \vect{D} _{\pm , \vect{Q}} ^{X *} \right| = D _{0} ^{X} / \sqrt{2}$ in Eq.(\ref{optical_matrix_element_LG_beam}) remains constant for all the $\vect{Q}$-exciton states coupled to the circularly polarized beam,  the optical matrix elements $\tilde{M} _{S , \vect{Q}} ^{\hat{\vect{\sigma}} ^{+} , 0 , 0} $  follows the same distribution over the $\vect{Q}$-space as that of the amplitude function $ \mathcal{A} _{0 0} \! \left( \vect{Q} \right)$ of the incident circularly polarized TL.
For reference, Figure.\ref{fig5}(a) re-presents the contour-plot of the $\left| \mathcal{A} _{0 0} \! \left( \vect{q}_{\parallel} \right) \right|^2$ of the fundamental LG beam in the $\vect{q}_{\parallel}$-space, schematically highlighted with the circular arrow lines representing the circular polarization, $\hat{\vect{\varepsilon}}=\hat{\vect{\sigma}} ^{+}$, carried by the incident LG beam.
With the beam waist of LG beam $W _{0} = 1.5$ $\mu$m, $\big| \mathcal{A} _{00} (\vect{q} _{\parallel}) \big| ^{2}  \propto \exp \big(- \! q _{\parallel} ^{2} \,  W _{0} ^{2} / 2 \big)$ spreads over a finite $\vect{q}_{\parallel}$-area at the scale of characteristic length $\sim 0.1 Q_c $, so does the $\big| \tilde{M} _{S , \vect{Q}} ^{\hat{\vect{\sigma}} ^{+} , 0 , 0} \big| ^{2} $ as seen in Fig.\ref{fig5}(c) and (e).

The optical matrix elements, $ \tilde{M} _{S , \vect{Q}} ^{\hat{\vect{\sigma}} ^{+} , 0 , 0} $, are known from Eq.(\ref{TID_wave_packet}) to determine the amplitudes of the finite-$\vect{Q}$ exciton components, $\tilde{g} _{S} ^{\hat{\vect{\sigma}} ^{+} , 0 , 0} (\vect{Q}) \propto  \tilde{M} _{S , \vect{Q}} ^{\hat{\vect{\sigma}} ^{+} , 0 , 0} $, in the TL-generated exciton wave packets. The magnitudes of $\tilde{g} _{+} ^{\hat{\vect{\sigma}} ^{+} , 0 , 0} (\vect{Q})$ [$\tilde{g} _{-} ^{\hat{\vect{\sigma}} ^{+} , 0 , 0} (\vect{Q})$] for the wave packet formed in the upper [lower] exciton band are represented by the sizes of filled circles placed on the $\vect{Q}$-exciton states of the upper [lower] band shown in Fig.\ref{fig5}(d) [(f)]. 
Fig.\ref{fig5}(d) and (f) show the finite-$\vect{Q}$ exciton components in the TL-excited wave packet that are distributed over the {\it upward} dispersions of the exciton bands. Accordingly, one can naturally infer the spectral {\it blue-shifts} of the exciton wave packets under the photo-excitation of the TL's in the higher-order modes, as observed by Ref.\cite{simbulan2021selective}.

\subsubsection{Shaping valley-exciton wave packets by using polarized TL's}

\begin{figure}[t]
\includegraphics[width=0.9\columnwidth]{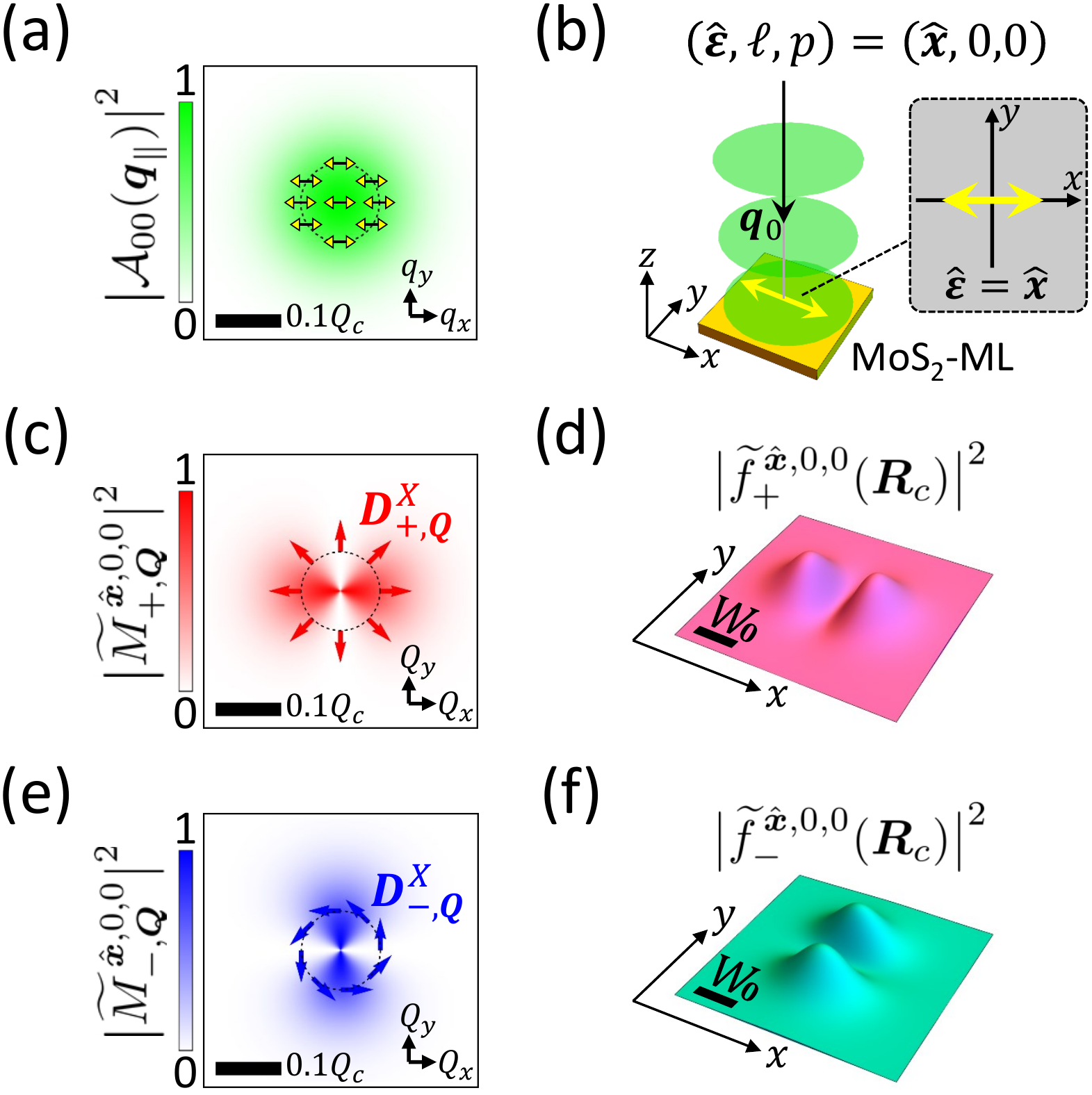}
\caption{(a) The square of the magnitude of the amplitude function, $\big| \mathcal{A} _{\ell p} (\vect{q} _{\parallel}) \big| ^{2}$, of normally incident linearly $\hat{\vect{x}}$-polarized LG beam in the fundamental mode, where the horizontal arrow lines indicate the linear polarization of the applied TL. (b) Schematics of a MoS$_{2}$ monolayer excited by the linearly polarized TL.  
(c) The square of the magnitude of the optical matrix elements in the $\vect{Q}$-space, $\left| \tilde{M} _{+ , \vect{Q}} ^{\hat{\vect{x}} , 0 , 0} \right| ^{2}$, and (d) the envelope functions in the $\vect{R} _{c}$-space, $\left| \tilde{f} _{+} ^{\hat{\vect{x}} , 0 , 0} (\vect{R} _{c}) \right| ^{2}$, of the finite-$\vect{Q}$  exciton states in the upper band photo-excited by the linearly polarized fundamental LG TL, where the black bar  in (d) represents the length of beam waist of the exciting TL, $W _{0} = 1.5$ $\mu$m. (e) and (f) present the results as (c) and (d), but for the transverse exciton states excited by the same polarized TL in the lower band.}
\label{fig6}
\end{figure}

Next, we turn to consider linearly polarized LG beams and examine the resulting exciton wave packets.
Figure.\ref{fig6}(c) and (e) show the square of magnitude of the complex optical matrix element $\big| \tilde{M}_{S , \vect{Q}} ^{\hat{\vect{x}} , 0 , 0} \big| ^{2}$ over the $\vect{Q}$-space for the {\it linearly} $\hat{\vect{x}}$-polarized LG beam in the fundamental $(0,0)$-mode. Differing from the cases of circularly polarized beam, the dipole-field coupling terms, $\hat{\vect{x}} \cdot \vect{D} _{- , \vect{Q}} ^{X *} = i D _{0} ^{X} \sin \phi _{\vect{Q}}$ and $\hat{\vect{x}} \cdot \vect{D} _{+ , \vect{Q}} ^{X *} = D _{0} ^{X} \cos \phi _{\vect{Q}}$, in Eq.(\ref{optical_matrix_element_LG_beam}) arising from the linearly polarized beam are no longer constant and highly depend on the orientation of the $\vect{Q}$-vector. With the $\phi_{\vect{Q}}$-dependence of the dipole-field coupling terms, the contour pattern of $\big| \tilde{M} _{S , \vect{Q}} ^{\hat{\vect{\varepsilon}} , \ell , p} \big| ^{2}$ over the $\vect{Q}$-space is reshaped to be highly anisotropic.

For the upper exciton band, as shown in Fig.\ref{fig6}(c), the $\hat{\vect{x}}$-polarized LG beam selectively couples the exciton states with the wavevectors surrounding the $Q _{x}$-axis. Correspondingly, the exciton wave packet photo-generated by the $\hat{\vect{x}}$-polarized LG beam for  the exciton upper band is more localized along the $\hat{\vect{x}}$-direction in the real space (that is parallel to the $\hat{\vect{x}}$-polarization of the exciting beam) and nearly vanishing around the $y$-axis in the real space, as described by $\big| \tilde{f} _{+} ^{\, \hat{\vect{x}} , 0 , 0} (\vect{R} _{c}) \big| ^{2}$ shown in Fig.\ref{fig6}(d).

By contrast, the photo-generated exciton wave packet in the lower band by the same $\hat{\vect{x}}$-polarized LG beam superpositions the finite momentum exciton states surrounding the $Q _{y}$-axis in the $\vect{Q}$-space and the resulting exciton wave packet turns out to be localized in the $\hat{\vect{y}}$-direction in the real space (that is perpendicular to the $\hat{\vect{x}}$-polarization of the exciting beam), as shown in Fig.\ref{fig6}(e) and (f).

The above studies show that the superposition of the finite-momentum exciton states in an exciton wave packet optically generated in the valley-split exciton bands by a TL can be engineered by the selection of the polarization of the exciting TL.
Besides, whereas the valley-mixed exciton bands split by only few meV are usually hardly resolved \cite{peng2019distinctive,simbulan2021selective,yu2014dirac,qiu2015nonanalyticity} and identified spectrally, the TL-excited exciton wave packets in the upper and lower bands present the completely distinctive $\vect{Q}$-distributions of the optical matrix elements, suggesting the easily distinguishable angle-dependent emitted lights from the upper- and lower-band wave packets (see the next section for detailed discussion).

\begin{figure}[t]
\includegraphics[width=0.9\columnwidth]{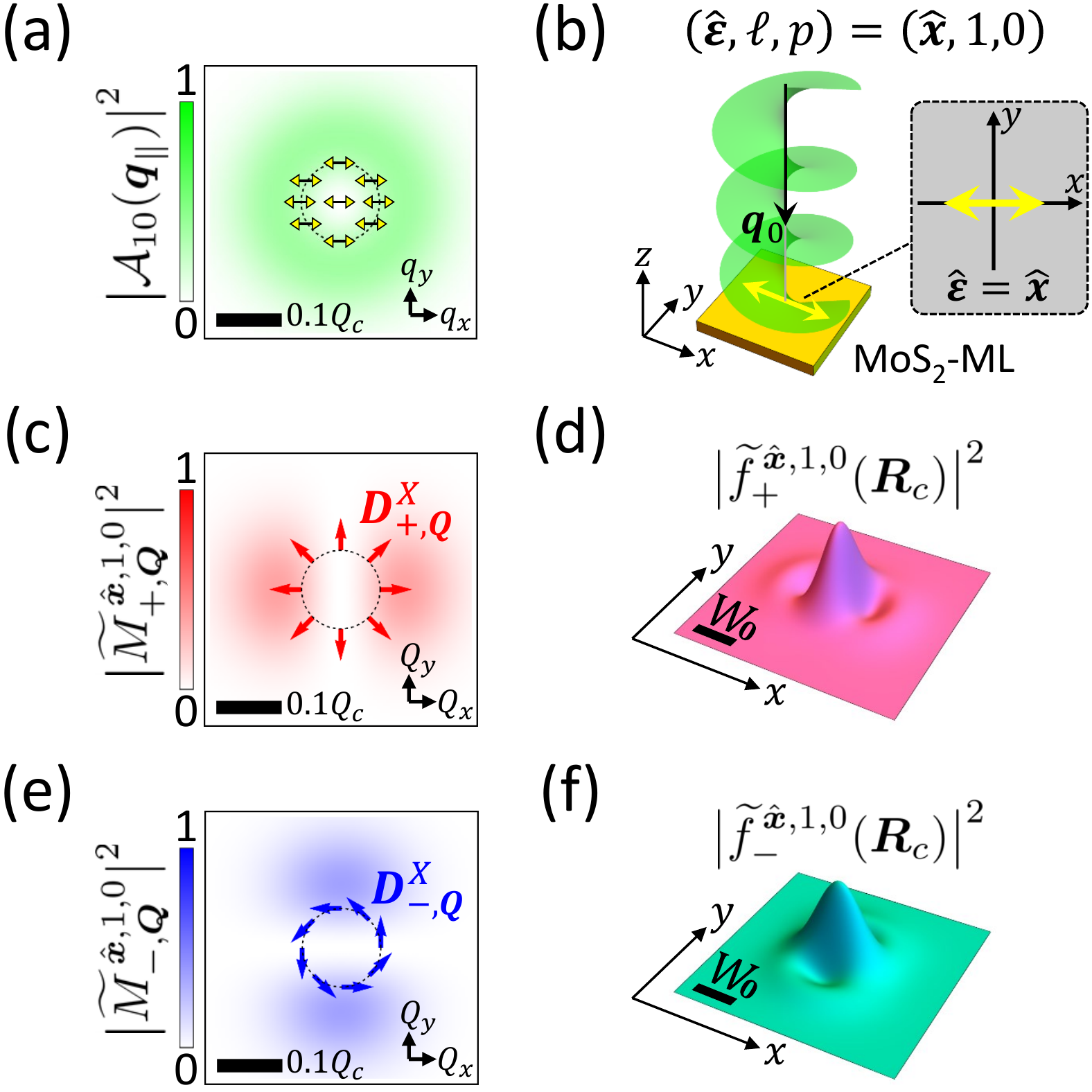}
\caption{(a) The square of the magnitude of the amplitude function, $\big| \mathcal{A} _{\ell p} (\vect{q} _{\parallel}) \big| ^{2}$, of the high-order LG beam carrying the finite OAM, $\ell=1$, with $\vect{q} _{0} = q _{0} \hat{\vect{z}}$ and the mode indices $( \hat{\vect{\varepsilon}} , \ell , p ) = ( \hat{\vect{x}} , 1 , 0 )$. (b) Schematics of a MoS$_{2}$ monolayer under the excitation of the linearly polarized TL with $( \hat{\vect{\varepsilon}} , \ell , p ) = ( \hat{\vect{x}} , 1 , 0 )$. (c) The square of the magnitude of the optical matrix elements in the $\vect{Q}$-space, $\left| \tilde{M} _{+ , \vect{Q}} ^{\hat{\vect{x}} , 0 , 0} \right| ^{2}$, and (d) the envelope functions in the $\vect{R} _{c}$-space, $\left| \tilde{f} _{+} ^{\hat{\vect{x}} , 0 , 0} (\vect{R} _{c}) \right| ^{2}$, of the finite-$\vect{Q}$ exciton states in the upper band photo-excited by the $\hat{\vect{x}}$-polarized TL with $\ell=1$, where the length of the black scale bar is, $W _{0} = 1.5$ $\mu$m, for reference. (e) and (f) present the results as (c) and (d), but for the TL-excited exciton states in the lower band.}
\label{fig7}
\end{figure}

\subsubsection{$\ell$- and $p$-dependent exciton wave packets}

After recognizing the shaping effect of light polarization (SAM of light) on TL-generated exciton wave packets, we proceed with the investigation of the photo-excitations in the TMD-ML's by the TL's carrying non-zero $\ell \neq 0$ (OAM of light).
Figure.\ref{fig7}(c) and (e) show the square of magnitudes of the $\vect{Q}$-dependent optical matrix elements $\big| \tilde{M} _{S = \pm , \vect{Q}} ^{\hat{\vect{x}} , 1 , 0} \big| ^{2}$ of the exciton wave packets in the upper and lower exciton band excited by the normally incident $\hat{\vect{x}}$-polarized LG beams in the mode of $(\ell, p)=(1,0) $, respectively. 
For reference of Fig.\ref{fig7}(c) and (e), Figure.\ref{fig7}(a) re-plots the amplitude function $\big| \mathcal{A}_{10} (\vect{q}_{\parallel}) \big| ^{2}$ featured with the ring-like contours in the $\vect{q}_{\parallel}$-space and highlighted with the horizontal arrow lines representing the $\hat{\vect{x}}$-polarization of the incident TL.  

From Eqs.(\ref{optical_matrix_element_LG_beam}) and (\ref{TID_wave_packet}), the amplitude function of the Fourier transform of the exciton wave packets induced by $\hat{\vect{x}}$-polarized TL's are given by $\tilde{g} _{S} ^{\hat{\vect{x}} , \ell , p} (\vect{Q}) = \frac{\tilde{M} _{S  , \vect{Q}} ^{\hat{\vect{x}} , \ell , p} }{\hbar \omega_0} \propto \mathcal{A} _{\ell p} (\vect{Q})  \big( \hat{\vect{x}} \cdot \vect{D} _{S , \vect{Q}} ^{X \, \ast} \big)$, which are determined by the product of the amplitude function of TL, $\mathcal{A} _{\ell p} (\vect{Q})$,  and the dipole coupling term, $\big( \hat{\vect{x}} \cdot \vect{D} _{S , \vect{Q}} ^{X \, \ast} \big)$. Accordingly, we obtain $\tilde{g} _{+} ^{\hat{\vect{x}} , \ell , p} \propto \mathcal{A} _{\ell p} (\vect{Q}) \cos \phi _{\vect{Q}}$ and $\tilde{g} _{-} ^{\hat{\vect{x}} , \ell , p} \propto \mathcal{A} _{\ell p} (\vect{Q}) \sin \phi _{\vect{Q}}$.
It turns out that, in Fig.\ref{fig7}(c)[(e)], the square of magnitude of the optical matrix elements of the upper [lower] exciton band, $\tilde{M} _{+, \vect{Q}} ^{\hat{\vect{x}} , \ell , 0} \propto \tilde{g} _{+} ^{\hat{\vect{x}} , \ell , 0} (\vect{Q}) \, \big[ \tilde{M} _{-, \vect{Q}} ^{\hat{\vect{x}} , \ell , 0} \propto \tilde{g} _{-} ^{\hat{\vect{x}} , \ell , 0} (\vect{Q}) \big]$ for $| \ell | = 1 $ becomes highly anisotropic and distributed mainly around the positive and negative $Q _{x}$-axes [$Q _{y}$-axis] in the $\vect{Q}$-plane. 

Correspondingly, Figure.\ref{fig7}(d) and (f) show the square of magnitude of the real-space envelope functions $\big| \tilde{f} _{S = \pm} ^{\hat{\vect{x}} , 1, 0} (\vect{R} _{c}) \big| ^{2}$ of the exciton wave packets photo-generated in the upper and lower exciton band, respectively, by the incident $(1,0)$ LG TL. As compared with the semi-ring-like wave packets excited by the fundamental LG beam with the nodes at the origin $\vect{R} _{c} = \vect{0}$  (see Fig.\ref{fig6}(d) and (f)), the wave packets excited by the LG beam with $\ell=1$ shown in Fig.\ref{fig7}(d) and (f) are nodeless at the origin position. According to Eq.(\ref{TID_wave_packet}), the envelope function of an exciton wave packet at the origin $\vect{R} _{c} = \vect{0}$ is given by $\tilde{f} _{\pm} ^{\hat{\vect{x}} , \ell , p} (\vect{R} _{c} = \vect{0}) \propto \int d ^{2} \vect{Q} \, \tilde{g} _{\pm} ^{\hat{\vect{x}} , \ell , p} (\vect{Q}) $, where $ \tilde{g} _{\pm} ^{\hat{\vect{x}}, \ell , p}  \propto \tilde{M} _{\pm} ^{\hat{\vect{x}} , \ell , p} (\vect{Q}) $. 
Following Eq.(\ref{optical_matrix_element_LG_beam}), one can show that $\tilde{g} _{\pm} ^{\hat{\vect{x}} , \ell , p} (\vect{Q}) \propto e ^{i \ell \phi _{\vect{Q}}} \big(e ^{i \phi _{\vect{Q}}} \pm e ^{-i \phi _{\vect{Q}}} \big)$, where the term $e ^{i \ell \phi _{\vect{Q}}}$ is arising from $\mathcal{A} _{\ell p} (\vect{Q})$ and the one $\big(e ^{i \phi _{\vect{Q}}} \pm e ^{-i \phi _{\vect{Q}}} \big)$ is from the dipole coupling term, $\big( \hat{\vect{x}} \cdot \vect{D} _{\pm , \vect{Q}} ^{X \, \ast} \big)$, of Eq.(\ref{optical_matrix_element_LG_beam}). Thus, one can derive that $\tilde{f} _{\pm} ^{\hat{\vect{x}} , \ell , p} (\vect{R} _{c} = \vect{0}) \propto \int _{0} ^{2 \pi} d \phi _{\vect{Q}} \ e ^{i \ell \phi _{\vect{Q}}} \big(e ^{i \phi _{\vect{Q}}} \pm e ^{-i \phi _{\vect{Q}}} \big) \propto \big( \delta _{\ell , -1} \pm \delta _{\ell , 1} \big)$. This shows that, under the linearly polarized TL excitation, the TL-excited exciton wave packets normally should have the nodes at $ \vect{R} _{c} =\vect{0}$, unless the exciting TL's carrying the OAM's, $\ell=\pm 1$. The exceptional hut-like wave packets nodeless at the origin results from the cancellation of the phase winding numbers, $n_{w}^{\ell}=\pm 1$, of optical OAM $\ell=\pm 1$ and those of the $\vect{Q}$-dependent dipoles of valley exciton, $n_{w}^X = \mp 1$, and manifest the interplay between the optical OAM and dipolar valley exciton.

\begin{figure}[t]
\includegraphics[width=0.9\columnwidth]{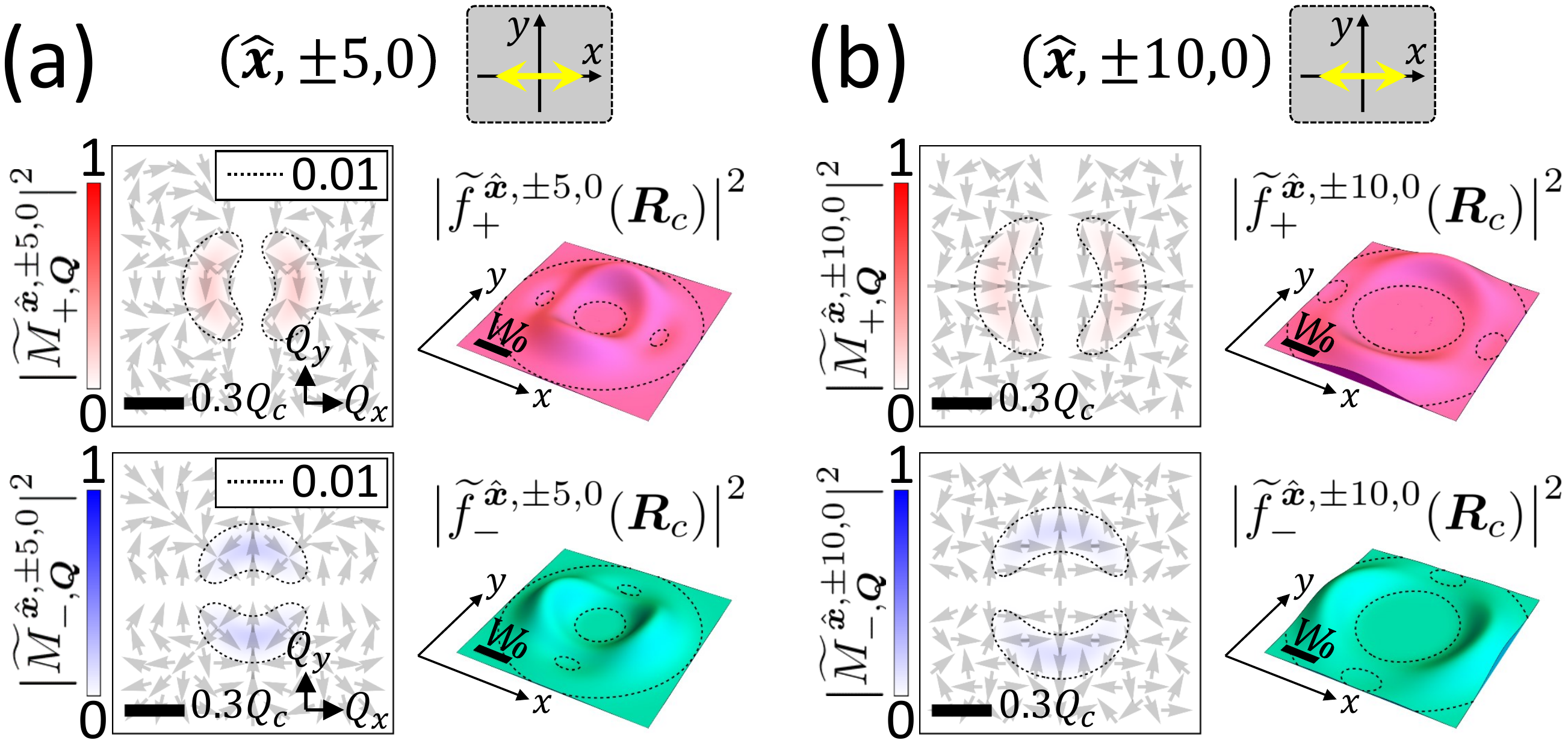}
\caption{The square of the magnitude of the optical matrix elements, $\big| \tilde{M} _{S , \vect{Q}} ^{\hat{\vect{\varepsilon}} , \ell , p} \big| ^{2}$, and the real-space envelope functions, $\big| \tilde{f} _{S , \vect{Q}} ^{\hat{\vect{\varepsilon}} , \ell , p} (\vect{R} _{c}) \big| ^{2}$, of the EWP's in a MoS$_2$-ML photo-generated by the higher-order $\hat{\vect{x}}$-polarized LG TL's with (a) $( \hat{\vect{\varepsilon}} , \ell , p ) = (\hat{\vect{x}} , \pm 5 , 0)$ and (b) $( \hat{\vect{\varepsilon}} , \ell , p ) = (\hat{\vect{x}} , \pm 10 , 0)$. The dotted contour lines in all plots have the equal value of 0.01. The gray arrows placed over the $\vect{Q}$-space are the vectors representing the complex values of the $\vect{Q}$-dependent optical matrix elements, $\tilde{M} _{S , \vect{Q}} ^{\hat{\vect{\varepsilon}} , \ell , p}$. }
\label{fig8}
\end{figure}

Further, let us examine the effects of the LG TL's with higher $\ell$'s. Figure.\ref{fig8} (a) and (b) show the $\vect{Q}$-dependent optical matrix elements and the real-space envelope functions of the exciton wave packets photo-generated by the linearly $\hat{\vect{x}}$-polarized LG TL's with $|\ell|=5$ and $10$, respectively.  
As shown in Fig.\ref{fig8}, the two semi-ring contour patterns of $\big| \tilde{M} _{\pm , \vect{Q}} ^{\hat{\vect{x}} , \ell, 0} \big| ^{2}$ with increasing $| \ell |$ are separated farther in the $\vect{Q}$-space and dominated by the exciton states with greater momenta. 
The contour patterns of $\big| \tilde{M} _{\pm , \vect{Q}} ^{\hat{\vect{x}} , \ell, 0} \big| ^{2}$ for the high $\ell$ are nearly vanishing over the extended reciprocal areas that are enlarged with increasing $\ell$. This implies the suppression of the vertical luminescences from the exciton wave packets photo-generated by the TL's in the higher $\ell$-modes.  
Correspondingly, the real-space exciton wave packets of the upper (lower) band photo-excited by the high-$\ell$ LG TL's are split into the shape of saddle featured with the nodes at $\vect{R} _{c} = \vect{0}$ and elongated along the positive and negative $y$-semi-axis ($x$-semi-axis) as shown in Fig.\ref{fig8}, consistent with the predicted nodes at $\vect{R} _{c} = \vect{0}$ for $\ell \neq \pm 1$ by our previous analysis.

\begin{figure}[t]
\includegraphics[width=0.9\columnwidth]{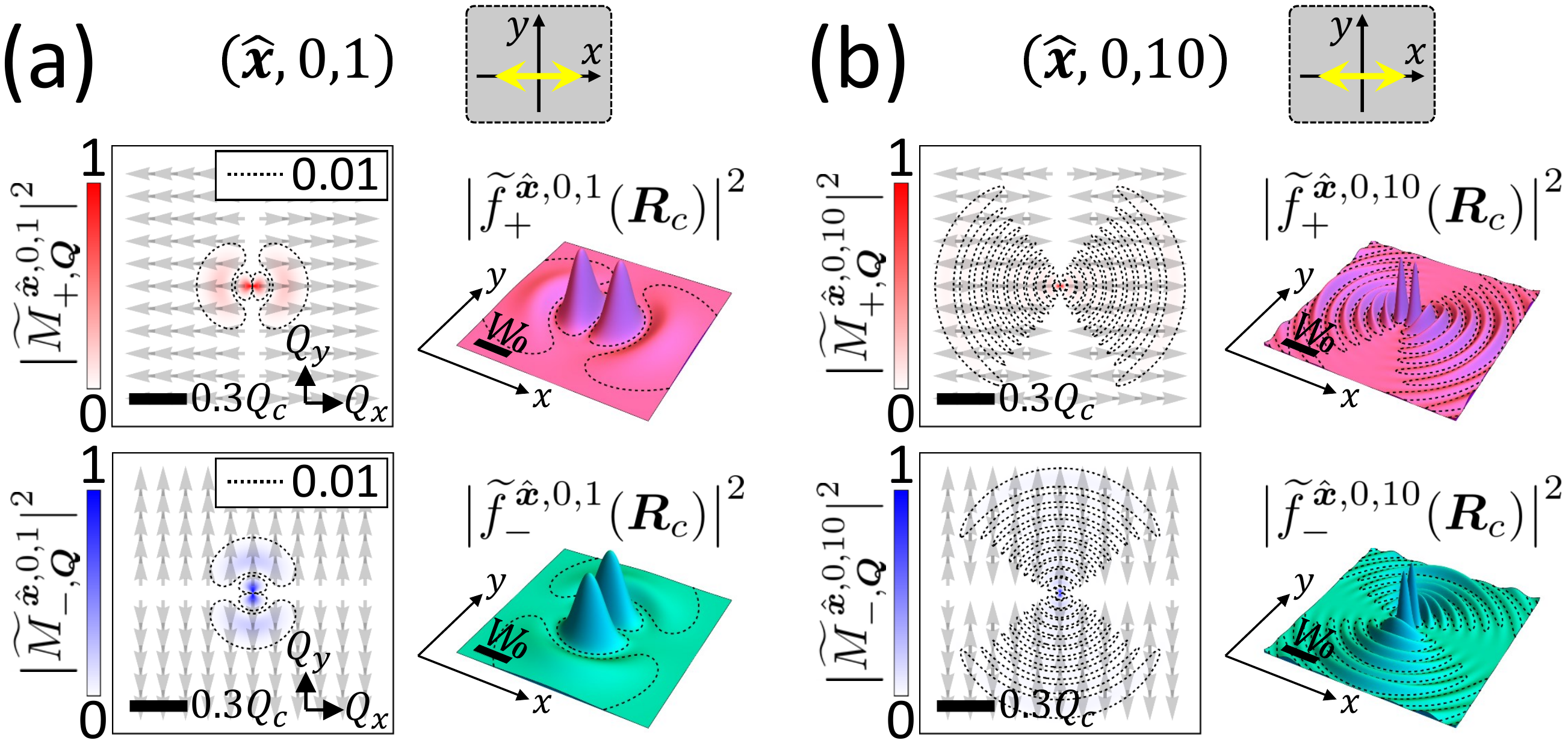}
\caption{The square of the magnitude of the optical matrix elements and envelope functions of the EWP's  photo-generated by the $\hat{\vect{x}}$-polarized LG TL's in the high-order radial modes with (a) $( \hat{\vect{\varepsilon}} , \ell , p ) = (\hat{\vect{x}} , 0 , 1)$ and (b) $( \hat{\vect{\varepsilon}} , \ell , p ) = (\hat{\vect{x}} , 0 , 10)$.}
\label{fig9}
\end{figure}

Besides the OAM's, the spatial structures of LG TL's are characterized also by the radial indices, $p$.
Figure.\ref{fig9} shows the amplitude functions of exciton wave packets, $\big| \tilde{M} _{\pm , \vect{Q}} ^{\hat{\vect{x}} , 0 , p} \big| ^{2}$, in $\vect{Q}$-space and the corresponding envelope functions, $\big| \tilde{f} _{\pm} ^{\hat{\vect{x}} , 0 , p} (\vect{R} _{c}) \big| ^{2}$, in real space resulting from the photo-excitations of normally incident $\hat{\vect{x}}$-polarized LG beams with $\ell=0$ and $p= 1, 10 $.
One notes that the $\vect{Q}$-dependent amplitude functions and the real-space envelop functions of the exciton wave packets excited by the LG TL's with $p\neq 0$ exhibit very different patterns from those excited by OAM TL's with $p=0$ shown by Figs.\ref{fig6}-\ref{fig8}.
Inheriting from the multiple-ring patterns of the amplitude function $\big| \mathcal{A} _{0 p} (\vect{Q}) \big| ^{2}$ of the high order LG TL's with $p\neq 0$ shown in Fig.\ref{fig4}, the optical matrix elements $\big| \tilde{M} _{\pm,\vect{Q}} ^{\hat{\vect{x}} , 0 , p} \big| ^{2}$ of the excited exciton wave packets in the upper (lower) band in Fig.\ref{fig9} are featured with the rippling structures along the $Q _{x}$-axis ($Q _{y}$-axis) that is aligned (perpendicular) to the direction of polarization. Similar rippling features are also retained in the real-space envelope functions of the exciton wave packets, as shown by the plots of  $\left| \tilde{f} _{\pm} ^{\hat{\vect{x}} , 0 , p} (\vect{R} _{c}) \right|^2$ in Fig.\ref{fig9}.

\subsection{Angle-resolved photo-luminescence spectra}

\begin{figure}
\centering
\includegraphics[width=0.9\columnwidth]{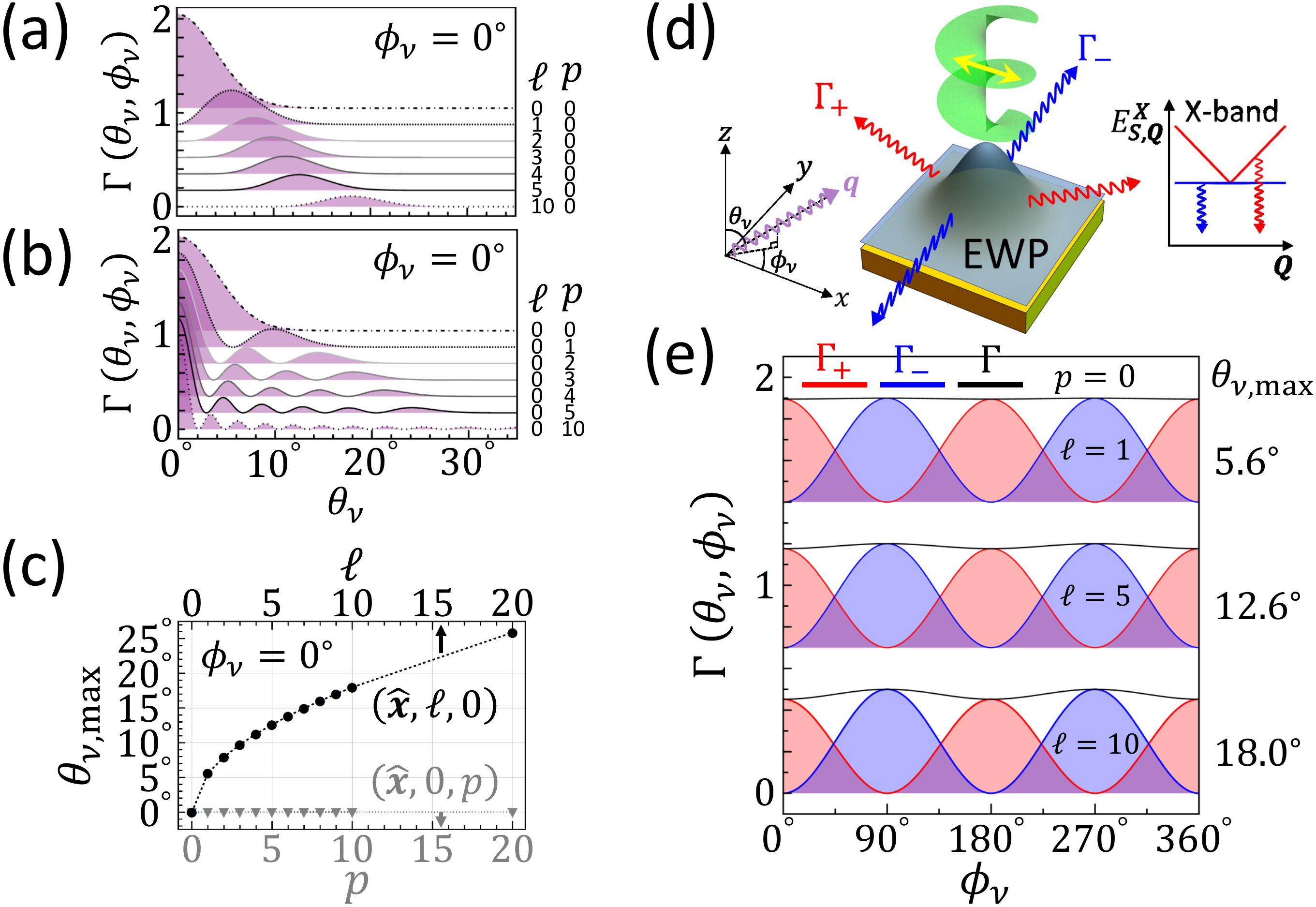}
\caption{The $\theta _{\nu}$-dependent emission rates, $\Gamma (\theta _{\nu} , \phi _{\nu}=0^\circ) = \sum_{S=+,-} \Gamma _{S} (\theta _{\nu} , \phi _{\nu})|_{\phi _{\nu}=0^\circ}$, of the directional PL's from the EWP's in a MoS$_{2}$ monolayer along the fixed $x-z$ plane ($\phi _{\nu}=0^\circ$) and with varied polar angles ($\theta _{\nu}$), photo-generated by the $\hat{\vect{x}}$-polarized TL's with (a) $(\hat{\vect{\varepsilon}} = \hat{\vect{x}} , \ell =0 , 1 , \cdots \! , 5, 10 , p = 0 )$ and  (b) $(\hat{\vect{\varepsilon}} = \hat{\vect{x}} , \ell = 0 , p =0 , 1 , \cdots \! , 5, 10 )$. From (a) and (b), (c) records the polar angles, $\theta _{\nu, \text{max}}$, of the maximum emission rates of the angle-resolved PL's from the EWP's photo-generated by the TL's in the various modes of $(\ell,0)$ and $(0,p)$.
The obvious $\ell$-dependence of $\theta _{\nu, \text{max}}$ indicates the OAM-encoded polar angles in the directional PL's from TL-excited EWP's.    (d) Schematics illustrating the polar, $\theta _{\nu}$, and azimuthal angle, $\phi _{\nu}$, of a directional PL with wavevector $\vect{q}$. (e) The $\phi _{\nu}$-dependence of $\Gamma_{S=\pm} (\theta_{\nu, \text{max}} , \phi _{\nu}) $ for the EWP's formed in the upper ($S=+$) and lower ($S=-$) exciton bands excited by the normal incident TL with $(\hat{\vect{\varepsilon}} = \hat{\vect{x}} , \ell =1, 5, 10 , p = 0 )$ at $\theta _{\nu} = \theta _{\nu , \text{max}}$. One notes that $\Gamma_{+}$ and $\Gamma_{-} $ follow the distinct and complementary $\phi _{\nu}$-dependences, allowing for selectively detecting the optical signals from the valley-split exciton bands according to the azimuthal angles. }
\label{fig10}
\end{figure}

The formation of exciton wave packets made by TL's should make impacts on the optical properties of the TL-excited TMD-ML's. In this work, we examine the angle-resolved photo-luminescences from TMD-ML's \cite{thompson2022valley} under TL-excitations, whose angle-dependent spectra reflect the $\vect{Q}$-dependent amplitude functions of the TL-generated exciton wave packets.

Under the incoherence condition, the {\it total} light emission rate of an exciton wave packet is determined by the Fermi's golden rule, summing up the emission rates of the plane-wave lights from individual $\vect{Q}$-exciton components in the wave packet, i.e. $\Gamma _{\text{spont}} = \sum _{\vect{q} , \lambda} \Gamma _{\vect{q}, \lambda}$ with
 $\Gamma _{\vect{q} , \lambda} = \frac{2 \pi}{\hbar} \big| \big\langle GS \big| \hat{\tilde{H}} _{I} ^{\vect{q} , \lambda \, \ast} (\vect{r}) \big| \Psi ^{X} \big\rangle \big| ^{2} \delta (\hbar \omega _{\vect{q}} - E _{S , \vect{0}} ^{X})$, where $\hat{\tilde{H}} _{I} ^{\vect{q} , \lambda \, \ast} (\vect{r})$ represents the second quantized operator of the light-matter interaction for plane-wave light with the wavevector $\vect{q}$ given by $\tilde{H} _{I} ^{\vect{q} , \lambda \, \ast} (\vect{r}) = \frac{|e|}{2 m _{0}} \big[ \vect{A} _{\vect{q}} ^{\hat{\vect{\varepsilon}} _{\vect{q} , \lambda}} (\vect{r}) \big] ^{\ast} \cdot \vect{p}$, $\lambda=1 (2)$ denotes the polarization basis of TM (TE) mode of emitted light, \cite{pedrotti2017introduction} $\vect{A} _{\vect{q}} ^{\hat{\vect{\varepsilon}} _{\vect{q} , \lambda}} (\vect{r}) = \hat{\vect{\varepsilon}} _{\vect{q} , \lambda} \, A _{0} \, e ^{i \vect{q} \cdot \vect{r}}$ is the vector potential of the plane-wave, and $\big| \Psi ^{X} \big\rangle$ is the exciton wave packet. For the purpose to examine the intrinsic optical properties of TL-induced EWP's, we consider  $\big| \Psi ^{X} \big\rangle$ simply given by Eq.(\ref{coherent_state}) as the initial state of the spontaneous PL and neglect the complex relaxation dynamics following the TL-excitation. Expressing the total spontaneous emission rate in the integral form of $\Gamma _{\text{spont}} = \frac{V}{(2 \pi) ^{3}} \int d ^{3} \vect{q} \sum _{\lambda} \Gamma _{\vect{q} , \lambda} \equiv \int d \theta _{\nu} \sin \theta _{\nu} \int d \phi _{\nu} \, \Gamma (\theta _{\nu} , \phi _{\nu})$, we can determine the rate of angle-resolved PL from an exciton wave packet induced by a $\hat{\vect{x}}$-polarized TL as
\begin{align}\label{angle_resolved}
  \Gamma ( \theta _{\nu} , \phi _{\nu} ) \propto \Big| \tilde{M} _{+ , \vect{q} _{c , \parallel}} ^{\hat{\vect{x}} , \ell , p} \Big| ^{2} \cos ^{2} \theta _{\nu} + \Big| \tilde{M} _{- , \vect{q} _{c , \parallel}} ^{\hat{\vect{x}} , \ell , p} \Big| ^{2} ,
\end{align}
where $V$ is the system volume, $\vect{q} _{c , \parallel} = Q _{c} \sin \theta _{\nu} \big( \cos \phi _{\nu} \, \hat{\vect{x}} + \sin \phi _{\nu} \, \hat{\vect{y}} \big)$ is the in-plane component of the emitted light wavevector $\vect{q} _{c} = \vect{q} _{c , \parallel} + Q _{c} \cos \theta _{\nu} \, \hat{\vect{z}}$ with $| \vect{q} _{c} | = Q _{c}$, and $\theta _{\nu}$ ($\phi _{\nu}$) is the polar (azimuthal) angle of the emitted light wavevector (see the schematics plotted in Fig.\ref{fig10}(d)). In the derivation of Eq.(\ref{angle_resolved}), we consider the vector of the polarization of TM wave, $\hat{\vect{\varepsilon}} _{\vect{q} _{c} , \lambda = 1} = \cos \theta _{\nu} \big( \cos \phi _{\nu} \, \hat{\vect{x}} + \sin \phi _{\nu} \, \hat{\vect{y}} \big) - \sin \theta _{\nu} \, \hat{\vect{z}}$, and that of TE one,  $\hat{\vect{\varepsilon}} _{\vect{q} _{c} , \lambda = 2} = - \sin \phi _{\nu} \, \hat{\vect{x}} + \cos \phi _{\nu} \, \hat{\vect{y}}$, that are coupled with the longitudinal and transverse exciton states of the upper and lower bands with $\vect{Q}=\vect{q} _{c , \parallel}$, respectively, and take the the approximation that $\omega _{S , \vect{Q}} = E _{S , \vect{Q}} ^{X} / \hbar  \approx  E _{S , \vect{0}} ^{X} / \hbar$ in the small $Q$-limit. The appearance of $\cos ^{2} \theta _{\nu}$ in the first term of $\Gamma (\theta _{\nu} , \phi _{\nu})$ in Eq.(\ref{angle_resolved}) arises from the projection of the polarization of TM light onto the  longitudinal dipoles of the upper-band exciton states, $\big| \hat{\vect{\varepsilon}} _{\vect{q} _{c} , 1} ^{\ast} \cdot \vect{D} _{+ , \vect{q} _{c , \parallel}} ^{X} \big| ^{2} \propto \cos ^{2} \theta _{\nu}$.

From Eq.(\ref{angle_resolved}), one can note that the angle-dependence of $\Gamma (\theta _{\nu} , \phi _{\nu})$ is correlated to the $\vect{q} _{c , \parallel}$-distribution of the optical matrix elements, $\tilde{M} _{\pm , \vect{Q}=\vect{q} _{c , \parallel}} ^{\hat{\vect{x}} , \ell , p}$. Taking the relation of $\tilde{M} _{\pm , \vect{q} _{c , \parallel}} ^{\hat{\vect{x}} , \ell , p} \propto \mathcal{A} _{\ell p} (\vect{q} _{c , \parallel}) \big( \hat{\vect{x}} \cdot \vect{D} _{\pm , \vect{q} _{c , \parallel}} ^{X \, \ast} \big)$ from Eq.(\ref{optical_matrix_element_LG_beam}), the angle-resolved PL from a TL-induced exciton wave packet in Eq.(\ref{angle_resolved}) composed of the contributions from the upper and lower exciton bands reads, $\Gamma (\theta _{\nu} , \phi _{\nu})= \Gamma_+ (\theta _{\nu} , \phi _{\nu}) + \Gamma_{-}(\theta _{\nu} , \phi _{\nu})$, where $\Gamma_+  \propto \big| \mathcal{A} _{\ell p} (\vect{q} _{c , \parallel}) \big| ^{2}  \cos ^{2} \phi _{\nu} \cos ^{2} \theta _{\nu}   $ and $\Gamma_-  \propto \big| \mathcal{A} _{\ell p} (\vect{q} _{c , \parallel}) \big| ^{2}  \sin ^{2} \phi _{\nu}$.

Accordingly, Figure~\ref{fig10}(a) and (b) show the $\theta _{\nu}$-dependence of the emission rate, $\Gamma (\theta _{\nu} , 0)$, of the angle-resolved PL's propagating in the $x-z$ plane ($\phi _{\nu}=0$) from the exciton wave packets in a TMD-ML photo-excited by the TL's with $\{\ell = 0, 1,2,..., p = 0\}$ and $\{ \ell = 0 , p = 0, 1,2,... \}$, respectively. Since $\Gamma_{-} (\theta _{\nu} , \phi _{\nu} = 0)\propto \sin ^2\left( \phi _{\nu} =0\right)=0$, the PL propagating in the $x-z$ plane is solely from the upper exciton band at the emission rate $\Gamma (\theta _{\nu} , 0) = \Gamma_{+} (\theta _{\nu} , 0)  \propto \big| \mathcal{A} _{\ell p} (\vect{q} _{c , \parallel}) \big| ^{2} \cos ^{2} \theta _{\nu} $. Hence, with increasing $\ell$, the direction of the highest PL is tilted from the $z$-axis with increasing $\theta_{\nu} \equiv \theta_{\nu,\text{max} }$ as shown in Fig.\ref{fig10}(a), following the similar $q _{x}$-dependence to that of $\big| \mathcal{A} _{\ell p} (\vect{q} _{c , \parallel}) \big| ^{2}$ as shown in Fig.\ref{fig3}(a). By contrast, as shown in Fig.\ref{fig10}(b), the angle-resolved PL with the maximum emission rate remains at $\theta_{\nu}=0$ with fixed $\ell=0$ and increasing $p=0, 1,2,...$, consistent with the the $q _{x}$-dependences of $\big| \mathcal{A} _{\ell p} (\vect{q} _{c , \parallel}) \big| ^{2}$ shown in Fig.\ref{fig4}(a). This suggests that angle-resolved PL measurements allow us better distinguish and identify the TL-generated exciton wave packets with different $\ell$. Figure.\ref{fig10}(c) shows how the polar angles, $\theta _{\nu, \text{max}}$, of the angle-resolved PL with the maximum emission rates of the TL-excited exciton wave packets depend on the OAM's (filled circles) and the radial indices (filled triangles) of the exciting TL's with $(\ell, p=0)$ and $(\ell=0, p)$, respectively. Apparently, the former shows the better angle-resolved spectra for varying $\ell$.

Finally, let us examine the $\phi _{\nu}$-dependences of the angle-resolved PL's from exciton wave packets in TMD-ML's excited by LG TL's.
Figure.\ref{fig10}(e) shows the emission rates of the directional PL's from the exciton wave packets in the upper and lower exciton bands excited by the TL's with OAM's, $\ell = 1, 5, 10$, over the full azimuthal angular range $\phi _{\nu}=\{0^\circ, 360^\circ\}$ with the fixed $\theta _{\nu}=\theta _{\nu, \text{max}}$.  One notes that, under the $\hat{\vect{x}}$-polarized TL-excitation, the directional PLs from the upper and lower exciton bands exhibit the distinctive and complementary $\phi _{\nu}$-dependences. As one sees in Fig.\ref{fig10}(e), the emission rate of the angle-resolved PL's from the upper-band wave packet oscillates with $\phi _{\nu}$, featured with the maximum rates at the angles $\phi _{\nu}=0^\circ , 180^\circ$, i.e. along the positive and negative $x$-axis in {\it parallel} to the $\hat{\vect{x}}$-polarization of the exciting TL's.
By contrast, the angle-resolved PL's from the lower-band wave packets show the maxima rates at the angles $\phi _{\nu}=90^\circ , 270^\circ$, i.e. along the positive and negative $y$-axis {\it perpendicular} to the $\hat{\vect{x}}$-polarization of the incident TL's. As a main finding of this work, we show that the PL's from specific valley-mixed exciton bands of TMD-ML's under polarized TL's are highly directional and dependent on the direction of the polarization of the applied TL's.
While the meV-split upper and lower exciton bands of a TMD-ML are normally hardly resolved spectrally, the angle-resolved PL spectroscopy on the TMD-ML's under the linearly polarized TL excitation is found to mimic an exciton multiplexer allowing for selecting and detecting the optical signatures of the valley-split exciton bands of TMD-ML's. \cite{laziifmmode2014scalable,katznelson2022bright}

\section{Conclusion}

In conclusion, we present a comprehensive theoretical investigation of the photo-generated valley excitons in TMD-ML's by Laguerre-Gaussian beams, one of the best known twisted lights carrying optical OAM, in addition to optical SAM. We show that a normally incident LG beam to a TMD-ML photo-generates spatially localized exciton wave packets constituted by the superposition of the finite-momentum exciton states, whose distributions over the exciton-momentum space are determined by the intriguing interplay between multiple photonic and excitonic degrees of freedom, including the OAM and SAM of light, and the valley pseudo-spin and center-of-mass motion of exciton.  This reveals the possibilities of using polarized TL's with engineered OAM to optically localize, shape and guide the wave packets of exciton, a key feature desired by the advanced optoelectronic applications and fundamental research that require charge-neutral excitons to be transported. The spatial structures of exciting TL's are shown directly to impact the angle-resolved photo-luminescence spectra of the TL-generated exciton wave packets. The polar angles of the directional PL's from the TMD-ML's under TL-excitations are increased by increasing the OAM of the exciting TL and serve for decoding the optical OAM transferred to excitons.   
Under the excitation of linearly polarized TL's, the photo-generated exciton wave packets formed in the upper and lower exciton bands yield the distinct and complementary azimuth-angle dependences, which mimic an exciton multiplexing set-up for selecting and detecting the optical signatures from the valley-split longitudinal and transverse exciton bands that are usually very hardly resolved spectrally.

\begin{acknowledgments}
This study is supported by the Ministry of Science and Technology, Taiwan, under contracts, MOST 109-2112-M-009 -018 -MY3, and by National Center for High-Performance Computing (NCHC), Taiwan. The authors are grateful to Ting-Hua Lu, Yann-Wen Lan and Shao-Yu Chen for fruitful discussions. 
\end{acknowledgments}

\bibliography{twisted_light_and_exciton_refs}{}






\end{document}